\documentclass[12pt]{article}
\usepackage[dvips]{color,graphicx}
\usepackage{hyperref}
\usepackage{amsmath,amssymb}
\usepackage{epsfig}
\input epsf.tex

\newcommand{\ls}{\ensuremath{l_s}} 

\def\p{\partial}

\newcommand{\tr}{\mathop{\rm Tr}}

\def\expec#1{\langle #1 \rangle}

\newcommand{\cF}{{\mathcal{F}}}

\newcommand{\cN}{{\mathcal{N}}}
\newcommand{\cO}{{\mathcal{O}}}

\newcommand{\bS}{{\mathbf{S}}}

\newcommand{\nn}{\nonumber}
\newcommand{\tret}{{t_{\mbox{\scriptsize ret}}}}

\newcommand{\ts}{{\tilde{s}}}
\newcommand{\tj}{{\tilde{j}}}
\newcommand{\tv}{{\tilde{v}}}
\newcommand{\ta}{{\tilde{a}}}
\newcommand{\tx}{{\tilde{x}}}
\newcommand{\ttau}{{\tilde{\tau}}}

\newcommand{\trFsq}{\tr F^2}

\newcommand{\vX}{\mbox{$\vec{X}$}}
\newcommand{\vx}{\mbox{$\vec{x}$}}
\newcommand{\Umax}{{U_{\mbox{\scriptsize max}}}}
\newcommand{\Umin}{{U_{\mbox{\scriptsize min}}}}
\newcommand{\Uminsq}{{U^2_{\mbox{\scriptsize min}}}}

\newcommand{\be}{\begin{equation}}
\newcommand{\ee}{\end{equation}}
\newcommand{\bea}{\begin{eqnarray}}
\newcommand{\eea}{\end{eqnarray}}

\begin{document}

\begin{titlepage}

\begin{flushright}
UTTG-12-11\\
TCC-014-11
\end{flushright}

\begin{center} \Large \bf The Gluonic Field of a Heavy Quark in\\
Conformal Field Theories at Strong Coupling
\end{center}

\begin{center}
Mariano Chernicoff$^{\star}$\footnote{mchernicoff@ub.edu},
Alberto G\"uijosa$^{\dagger}$\footnote{alberto@nucleares.unam.mx}
and Juan F.~Pedraza$^{\natural}$\footnote{jpedraza@physics.utexas.edu}

\vspace{0.2cm}
$^{\star}$
Departament de F{\'\i}sica Fonamental and Institut de Ci\`encies del Cosmos,\\
Universitat de Barcelona,\\
Marti i Franqu\`es 1, E-08028 Barcelona, Spain\\
 \vspace{0.2cm}
$^{\dagger}\,$Departamento de F\'{\i}sica de Altas Energ\'{\i}as, Instituto de Ciencias Nucleares, \\ Universidad Nacional Aut\'onoma de
M\'exico,\\ Apartado Postal 70-543, M\'exico D.F. 04510, M\'exico\\
 \vspace{0.2cm}
$^{\natural}\,$Theory Group, Department of Physics and Texas Cosmology Center, \\
University of Texas, 1 University Station C1608, Austin, TX 78712, USA\\
\vspace{0.2cm}
\end{center}

\begin{center}
{\bf Abstract}
\end{center}
\noindent
We determine the gluonic field configuration sourced by a heavy quark undergoing arbitrary motion in $\cN=4$ super-Yang-Mills at strong coupling and large number of colors. More specifically, we compute the expectation value of the operator $\tr[F^2+\ldots]$ in the presence of such a quark, by means of the AdS/CFT correspondence. Our results for this observable show that signals propagate without temporal broadening, just as was found for the expectation value of the energy density in  recent work by Hatta \emph{et al.} We attempt to shed some additional light on the origin of this feature, and propose a different interpretation for its physical significance. As an application of our general results, we examine $\expec{\tr[F^2+\ldots]}$ when the quark undergoes oscillatory motion, uniform circular motion, and uniform acceleration. Via the AdS/CFT correspondence, all of our results are pertinent to any conformal field theory in $3+1$ dimensions with a dual gravity formulation.

\vspace{0.2in}
\smallskip
\end{titlepage}
\setcounter{footnote}{0}

\tableofcontents

\section{Introduction and Summary}

\subsection{Motivation}

When a charge moves, it produces a propagating disturbance in the associated gauge field. The problem of determining the spacetime profile of this disturbance for an arbitrary charge trajectory was solved long ago for classical electrodynamics \cite{jackson}, but, under various guises, remains of significant interest today in the context of quantum non-Abelian gauge theories. In recent years, gauge/gravity duality \cite{malda,gkpw,magoo} has given us a useful handle on this and many other problems for a varied class of gauge theories in the previously inaccessible regime of strong coupling, via a drastic and surprising rewriting in terms of string-theoretic degrees of freedom living on a curved higher-dimensional geometry.

Among the known examples of the duality, the best understood subclass is that of conformal field theories (CFTs), where the relevant curved geometry is asymptotically anti-de Sitter (AdS). In this paper we will use this AdS/CFT correspondence to study the  propagation of disturbances in the gluonic field produced by a moving heavy quark in a strongly-coupled conformal gauge theory, in the limit where the number of colors is large. For concreteness, we will phrase our analysis in terms of $\cN=4$ super-Yang-Mills (SYM), even though the results we will obtain are equally relevant to other CFTs in $3+1$ dimensions, can  easily be extended to CFTs in other dimensions, and might also be expected to apply at a qualitative level in some of the non-conformal examples of the gauge/gravity correspondence.

The AdS/CFT correspondence states that a heavy quark moving in the vacuum of $\cN=4$ SYM is dual to a string moving on a pure AdS$_5$ geometry. More precisely, the quark corresponds to the endpoint of a string, whose body codifies the profile of the non-Abelian (near and radiation) fields sourced by the quark. When the quark/endpoint moves, it generally produces a wave running along the body of the string, which corresponds to the wave in the gluonic field whose profile we are interested in determining.

The translation between string and gauge theory disturbances was first explored in \cite{cg}, which used tools developed in \cite{gkpw,dkk} to study the dilaton waves given off by
fluctuations on an otherwise static, radial string in AdS$_5$, and infer from them the profile of the dual gluonic field observable
$\expec{\tr F^2(x)}$ in the presence of an oscillating quark. Under the assumption that these oscillations are small, the authors of \cite{cg} treated the string dynamics in a linearized approximation. Their results painted an interesting picture of wave propagation in $\cN=4$ SYM: in contrast with the standard Lienard-Wiechert story of the classical Abelian case, where signals propagate strictly at the speed of light, the waves in $\expec{\tr F^2(x)}$ were found to display significant temporal broadening, just as one would expect given that points on the non-Abelian field arbitrarily far from the quark can themselves reradiate.

\begin{figure}[htb]
\centerline{\epsfxsize=9cm \epsfbox{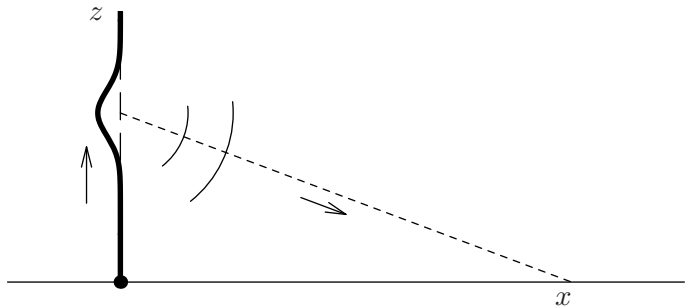}}
\caption{\small Schematic representation of emission of dilaton waves by a string in AdS. The radial direction  of AdS is denoted by $z$ and runs vertically, with the AdS boundary ($z=0$) located at the bottom of the figure. Any given point $x$ on the boundary receives a dilaton wave
from each point along the string. As a result, the gauge theory wave
is a superposition of components with all possible time delays. }
\label{signal}
\begin{picture}(0,0)
 \put(104,35){\small $x$}
 \put(42,73){\small $z$}
 \end{picture}
\end{figure}

As shown in Fig.~1, this feature emerges naturally in the gravity side of the correspondence, because motion of the string endpoint generates waves that move up along the body of the string, and each point on the string then emits a dilaton wave that travels back down to the observation point $x$ on the AdS boundary, where, via the AdS/CFT recipe for correlation functions \cite{gkpw}, the value of
$\expec{\tr F^2(x)}$ is deduced by assembling together all such contributions. The dilaton waves originating from points on the string that are further away from the AdS boundary give rise to components of the SYM wave that have a larger time delay. This makes perfect physical sense, because, through the UV/IR connection \cite{uvir}, such points are known to be dual to regions of the gluonic field further away from the quark. So, even when treating the string waves at the linearized level, the very fact that we are dealing with a string leads to gauge theory disturbances that display the nonlinear propagation expected from the non-Abelian character of the strongly-coupled SYM fields.

 The simple observable $\expec{\tr F^2(x)}$ that was the focus of \cite{cg} served well to exhibit the main features of propagating disturbances in the gluonic field, but
did not allow a definite identification of waves with the characteristic $1/|\vec{x}|^2$ falloff associated with
radiation, i.e., contributions that transport energy to infinity. Of course, in the case of an accelerating quark, fluctuations in the near-fields are fully expected to be accompanied by radiation proper. The unambiguous detection of the latter calls for examination of the SYM energy-momentum tensor, $\expec{T_{\mu\nu}(x)}$, which in the gravity side of the correspondence requires a determination of the gravitational waves emitted by the string. Indeed, some time after \cite{cg}, the $1/|\vec{x}|^2$ falloff was  established in the same setup through a
calculation of the (time-averaged) energy density, $\expec{T_{00}(x)}$ \cite{mo}.

 In more recent years, motivated by potential contact with the phenomenology of the quark-gluon plasma \cite{qgprev}, analyses of both  $\expec{\tr F^2(x)}$ and $\expec{T_{\mu\nu}(x)}$ have been carried out in the case of a heavy quark moving at constant velocity through a thermal plasma (where an exact solution for the corresponding string embedding is available \cite{hkkky,gubser}), in a large body of work that includes \cite{gluonicprofile} and has been reviewed in \cite{gluonicprofilerev}. The results are again compatible with the expected nonlinear dynamics of the (in this case, finite-temperature) SYM medium.

Given these antecedents, it came as a surprise when, back at zero temperature, additional calculations  going beyond the linearized string approximation found, first for special cases \cite{liusynchrotron,iancu1} and then for an arbitrary quark trajectory \cite{iancu2}, that the ensuing energy density $\expec{T_{00}(x)}$ displays no temporal broadening, and is in fact as sharply localized in spacetime as the corresponding classical profile. As shown in \cite{iancu1,iancu2}, the fact that disturbances in $\expec{T_{00}(x)}$ propagate strictly at the speed of light does not actually conflict with the UV/IR argument presented three paragraphs above, because the full gluonic profile can be understood as arising purely from a contribution of the string endpoint, and this is the reason why it features a single time delay.

Our work was motivated by the tension between these two sets of results for the gluonic fields in vacuum. How can it be that the $\expec{\tr F^2(x)}$ profile obtained in \cite{cg} displays significant temporal broadening, while the $\expec{T_{00}(x)}$ pattern deduced in \cite{iancu2} does not? A natural strategy to further explore and attempt to resolve this tension is to carry out the $\expec{\tr F^2(x)}$ calculation beyond the linearized string approximation, and for an arbitrary quark trajectory, to put it on a par with the computation in \cite{iancu2}. This is what we set out to do in this paper.

\subsection{Outline and main results}

We begin in Section \ref{recipesec} by presenting the ingredients and recipe for our calculation.  The desired one-point function of $\tr F^2(x)$ (or, more precisely, of the Lagrangian density operator (\ref{o})), in the presence of a moving quark, must be extracted  via (\ref{trfsq}) from the leading near-boundary behavior of the dilaton field (\ref{dilsol}) sourced by the string profile (\ref{mikhsolzm}). The latter is the unique string embedding  codifying the gluonic fields generated by a quark undergoing arbitrary motion, under the assumption that such fields are \emph{retarded} or \emph{purely outgoing}, i.e., they propagate outwards from the quark to infinity. This solution was obtained in \cite{mikhailov} for an infinitely heavy quark  and generalized  in \cite{dragtime,lorentzdirac,damping} to the case of a quark that has a finite mass, and, consequently, a finite size (i.e., Compton wavelength) $z_m$ given by (\ref{zm}).

The actual computation is carried out in Section \ref{arbitrarysec}. A key feature is that, when expressed in terms of appropriately geometric variables (the quark proper time $\tau$ and the invariant AdS distance (\ref{U})), a total derivative is found to appear in the worldsheet integrand (\ref{dilsolmikh2}), allowing one of the two integrals to be done trivially. This feature results from a nontrivial cancelation (noted below (\ref{Umikh})) that in turn originates from the retarded structure of the string embedding (\ref{mikhsol}). An analogous cancelation was found in the energy density computation in \cite{iancu2}, arising there from an interplay between the various components of the string energy-momentum tensor. Since our analysis involves only scalar quantities, it becomes clear that the cancelations in question are not intrinsically tied to the tensorial nature of the  $\expec{T_{00}(x)}$ derivation.

Our final result for the gluonic profile $\expec{\tr F^2(x)}$ in the case of a quark with finite mass is somewhat involved: it is written in (\ref{trfsqfinal}) in terms of auxiliary (tilde) variables whose explicit dependence on the quark's velocity, acceleration, jerk and snap, as well as on the external force (and the first and second derivatives thereof) it is subjected to, is given in (\ref{atilde})-(\ref{xtilde}), (\ref{jtilde})-(\ref{stilde}). It is easy to check that, when the quark is static, our general expression correctly reproduces the known \cite{martinfsq} finite-mass  result (\ref{trfsqstatic2}). For arbitrary motion, \emph{the gluonic field at any given observation point is found to depend only on dynamical data evaluated at a single retarded time along the quark trajectory}, specified by (\ref{tau0}) or, in non-covariant form, (\ref{tret}). This is the same surprising characteristic demonstrated for the radiation component of the field in \cite{iancu1,iancu2}.

In Section \ref{examplessec}, we apply the general result (\ref{trfsqfinal}) to three specific examples. The first is the oscillating quark that was the focus of \cite{cg} and the main motivation for our work. We start by performing a numerical analysis of the case (not covered by \cite{cg}) where the quark has a finite mass, and display the resulting gluonic profile in Fig.~4. Moving on to the case of infinite mass, we show explicitly that the integrated expression obtained in \cite{cg}, which manifestly displays the gluonic field as a sum of contributions with all possible time delays, actually gives the same result (within its range of validity) as the linearized and infinitely-massive version of our final formula (\ref{trfsqfinal}) for the gluonic profile, which incorporates a single time delay. This proves that, despite appearances, there is in fact no conflict between \cite{cg} and \cite{iancu1,iancu2}: it is just that the choice of worldsheet coordinates and the lack of an exact solution for the string embedding prevented the authors of \cite{cg} from being able to carry out the integral explicitly to find the \emph{net} retardation pattern.

Our second example is  uniform circular motion, studied in Section \ref{circularsubsec} to complement the energy density analysis of \cite{liusynchrotron}. Our third and final example, examined in Section \ref{accelsubsec}, is uniform acceleration, for which \cite{accelembedding,brownianunruh,brownianunruh2}
had already made some interesting physical inferences based on the structure of the string embedding.

In our final Section \ref{discussionsec} we go back to our result for arbitrary quark motion, and discuss its physical implications. We consider first the case of an infinitely heavy quark, which was the only one contemplated by the $\expec{\tr F^2(x)}$ computation in \cite{cg} and the $\expec{T_{00}(x)}$ calculations in \cite{liusynchrotron,iancu1,iancu2}.
In this limit, the quark becomes pointlike ($z_m\to 0$), and our result simplifies drastically, taking the form (\ref{trfsqpointlike}). For an arbitrary spacetime trajectory of the quark, this is manifestly just the boosted Coulomb profile associated with a uniformly translating quark with the retarded position and velocity inferred by projecting back along the past lightcone of the observation point. The full  profile agrees (up to an overall constant) with the corresponding Lienard-Wiechert result in classical electrodynamics \cite{jackson}. The observable $\expec{\tr F^2(x)}$ determined in this paper is thus seen to yield an image of the gluonic profile for a pointlike quark that is in complete consonance with the one obtained via $\expec{T_{00}(x)}$ in \cite{iancu2}, with the main difference being that, just like in electrodynamics, the former is sensitive only to disturbances in the near field of the quark, whereas the latter also explicitly incorporates radiation. Both here and in \cite{iancu2} signals in the gluonic field are found to propagate strictly at the speed of light (as indicated below (\ref{trfsqpointlike})), with no temporal (or, equivalently, radial) broadening.

Our more general result (\ref{trfsqfinal}), relevant for a quark with finite mass, differs from (\ref{trfsqpointlike}) and from the  energy density obtained in \cite{iancu2} in two ways. First, the retarded quark data are read off at a source point that is related to the observation point through the \emph{timelike} interval (\ref{tau0}). In other words, signals are found to propagate at a (variable) subluminal speed, which as seen in (\ref{vsignal}) can in fact be arbitrarily small. Second, the field profile depends on more data than just the position and velocity of the quark. As explained in Section \ref{discussionsec}, both of these features are naturally associated with the fact that the quark is no longer pointlike.

Even with these differences, it is still true that in our general result (\ref{trfsqfinal}) the field at a given observation point is controlled by data at a single retarded event, as in \cite{iancu1,iancu2}. The discrepancy between this pattern and the one reported in \cite{cg}  was discussed in \cite{iancu1} in terms of a fundamental distinction between the quark's near field (which is all that is visible in $\expec{\tr F^2(x)}$) and its radiation field (which dominates $\expec{T_{00}(x)}$ at long distances), but our results show that there is in fact no such distinction, because both components display exactly the same unbroadened propagation. The reason on the gravity side is the same in both cases: the fact that (in accord with the UV/IR connection) the result can be expressed in terms of a contribution arising purely from the string endpoint. Moreover, we know from Section \ref{oscillatesubsec} that the discrepancy between \cite{cg} and \cite{iancu1,iancu2} is only apparent, and that, in spite of the fact that the gluonic field emerges as a superposition of contributions with all possible time delays, the \emph{net} result evidences only the smallest of these delays.

In \cite{iancu1,iancu2} it was argued that the lack of broadening is unphysical and points to a deficiency of the AdS/CFT calculation. The argument visualizes the radiation process in terms of emission of gluons that are themselves able to reemit, and would be expected to do so profusely at strong coupling. This would bring into play large quantities of off-shell gluons, which would naturally be expected to propagate at subluminal speeds and therefore to lead inevitably to temporal/radial broadening of the emitted field. This `parton branching' picture is essentially a perturbative rephrasing of the discussion in \cite{cg} (recalled above) of non-Abelian reemission by the gluonic field at all different length scales, and has been shown to be consistent with several other AdS/CFT results \cite{branching}.
Given the apparent incompatibility of this picture with the lack of broadening, the authors of \cite{iancu1,iancu2} went on to suggest that, contrary to widespread belief, the supergravity approximation to physics on the gravity side does not capture the full quantum dynamics of the large $\lambda$ and large $N_c$ limit, and tried to identify the missing element as arising from longitudinal fluctuations in a heuristic lightcone gauge calculation.

As explained in Section \ref{discussionsec}, even though we find in this paper that $\expec{\tr F^2(x)}$ displays exactly the same retardation pattern as $\expec{T_{00}(x)}$, we do not subscribe to the point of view of \cite{iancu1,iancu2}. As seen in the calculation for arbitrary quark trajectory in Section \ref{arbitrarysec} (as well as in \cite{iancu2}), and also  when we make contact in Section \ref{oscillatesubsec} between \cite{cg} and our general result (\ref{trfsqfinal}),
the supergravity description assembles the gluonic field precisely in the physically expected manner, by summing over contributions reradiated from all possible length scales (as depicted schematically in Fig.~1), associated with all possible time delays. It is therefore not true that the absence of net temporal broadening in the final AdS/CFT result implies that the supergravity approximation is somehow leaving out the expected non-Abelian rescattering. Rather, we interpret the no-broadening result as a \emph{prediction} of the AdS/CFT correspondence for the \emph{net} pattern of propagation in the CFT at strong coupling and with a large number of colors.

 We stress in particular that the appearance in (\ref{dilsolmikh2}) of a total derivative in the worldsheet integrand, which enables us to present the resulting dilaton field as a pure endpoint contribution, does not mean that points on the body of the string do not contribute, but only that their aggregated contribution can be reexpressed in terms of the behavior at the edge of the integration region. The same applies then on the SYM side: the fact that the
 gluonic field at the observation point can be reported purely in terms of the behavior of the quark at a single retarded time does not indicate that only that instant contributes, but only that the cumulative effect of summing over the contributions from all previous emission events (arising via non-Abelian rescattering at all possible length scales) can be reexpressed in terms of the aforementioned behavior.
This is explicitly shown by our calculations, by those of \cite{liusynchrotron,iancu1,iancu2}, and also by a recent surprising reformulation of the latter as a superposition of gravitational shock waves emitted by each point along the string \cite{veronika}.

It is important to keep in mind that the form of our result depends crucially on the retarded structure of the worldsheet embedding (\ref{mikhsolzm}), which in turn follows from the assumption of a purely outgoing condition for the gluonic field generated by the quark. For any choice other than the purely outgoing (or purely ingoing) SYM configuration, we would expect not to obtain a total derivative on the worldsheet, and this would then lead to a final gluonic profile showing finite temporal/radial broadening. At least in retrospect, it is natural for the purely outgoing condition to impose a restriction on the overall retardation pattern of the field, because arbitrary reradiation from all points on the non-Abelian medium would result in wave scattering back towards the quark.  That the final gluonic profile at any given observation point, which receives contributions from the entire quark trajectory, can be reexpressed in terms of data at a single retarded time, is also not surprising \emph{per se}. At the calculational level, that happens every time we evaluate an integral in terms of the behavior at the integration endpoint. What is remarkable about the  restriction predicted by AdS/CFT is that it involves only a \emph{finite number} of quark data at the relevant instant.

In short, then, we believe our results resolve the apparent conflict between \cite{cg} and \cite{iancu1,iancu2}, and support an interpretation of the lack of temporal broadening that does not challenge the ability of the supergravity approximation of AdS/CFT to capture the full dynamics in the strong-coupling limit.

\section{Ingredients of the Computation}\label{recipesec}

We are interested in studying the gluonic field sourced by a heavy quark in a strongly-coupled gauge theory. Gauge/gravity duality grants us access to many different setups, and in particular, to conformal field theories (CFTs) in any dimension, but for concreteness we will focus on  $\cN=4$ super-Yang-Mills (SYM) with gauge group $SU(N_c)$. This is a conformally invariant theory with a vector field, 6 real scalars and 4 Weyl fermions, all in the adjoint representation of the gauge group.
The AdS/CFT correspondence \cite{malda} asserts that this theory, on $(3+1)$-dimensional Minkowski spacetime, is fully equivalent to Type IIB string theory on the Poincar\'e patch of the
AdS$_5\times\bS^5$ geometry,
\begin{equation}\label{metric}
ds^2=G_{mn}dx^m dx^n={R^2\over z^2}\left(
-dt^2+d\vec{x}^2+dz^2 \right)+R^2 d\Omega_5^2~,
\end{equation}
(with a constant dilaton and $N_c$ units of Ramond-Ramond
five-form flux through the five-sphere). The radius of curvature $R$ is related to the SYM 't Hooft coupling $\lambda\equiv g_{YM}^2 N_c$ through
\begin{equation}\label{lambda}
 \lambda={R^4\over \ls^4}~,\nonumber
\end{equation}
 where $\ls$ denotes the string length.
In more detail, the state of IIB string theory described by (\ref{metric}) corresponds to the (symmetry-preserving) vacuum of the gauge theory, and the closed
string sector describing small or large fluctuations on top of it encodes the gluonic ($+$
adjoint scalar and fermionic) physics.
The coordinates $x^{\mu}\equiv(t,\vec{x})$  parallel to the AdS boundary $z=0$  are
directly identified with the gauge theory spacetime coordinates, the
radial direction $z$ is mapped to a variable length (or, equivalently, inverse energy) scale in SYM
\cite{uvir},  and the five-sphere coordinates are associated with the global $SU(4)$ internal (R-) symmetry of SYM. These angular coordinates will play no role in our analysis, so our results will hold equally well in the more general case where the $\bS^5$ is replaced by a different compact five-dimensional space $\mathbf{X}_5$, which corresponds to replacing $\cN=4$ SYM with a different $(3+1)$-dimensional CFT.

The introduction of an open string sector associated with a
stack of $N_f$ D7-branes in the geometry
(\ref{metric}) is equivalent \cite{kk} to the addition, on the gauge theory side,  of $N_f$ hypermultiplets
(each composed of a Dirac fermion and 2 complex scalars) in the \emph{fundamental}
representation of the $SU(N_c)$ gauge group, that we will refer to as `quarks'.
 For $N_f\ll N_c$,
 we are allowed to neglect the backreaction of the D7-branes on the geometry;\footnote{More precisely, the relevant condition is $\lambda N_f \ll N_c$ \cite{unquenched}.}
 in the gauge theory this corresponds to working in a `quenched' approximation
that ignores quark loops. The D7-branes
cover the four gauge theory directions $x^{\mu}$, and extend
along the radial AdS direction up from the boundary at $z=0$ to a
position where they `end' (meaning that the $\bS^3\subset\bS^5$
that they are wrapped on shrinks down to zero size), whose
location $z=z_m$ is related to the mass $m$ of the quarks through
\begin{equation}\label{zm}
z_m={\sqrt{\lambda}\over 2\pi m}~.
\end{equation}

An isolated quark is dual to an open string that extends radially from the location $z=z_m$ on the D7-branes to the horizon of the Poincar\'e patch,  $z\to\infty$. We will describe the dynamics
 of our string in first-quantized language, and since we take it to be heavy, we are allowed to treat it semiclassically. In gauge theory language,
then, we are coupling a first-quantized quark to the gluonic ($+$ other SYM) field(s), and
carrying out the full path integral over the strongly-coupled field(s) (the result of which is
codified by the AdS spacetime), but treating the path integral over the quark trajectory
$\vec{x}(t)$ in a saddle-point approximation.

In the nonperturbative framework provided to us by the AdS/CFT correspondence, a quark with finite mass ($z_m>0$) is automatically not `bare' but `composite' or `dressed'.  This can be inferred, for instance, from the expectation value of the gluonic field surrounding a static quark,
as in \cite{martinfsq} and the calculations we will perform in the next section, or from the deformed nature of the quark's dispersion relation \cite{dragtime,lorentzdirac,damping}.
The characteristic thickness of the `gluonic cloud' surrounding the quark
 is given by $z_m$, which  is thus the analog of the Compton wavelength
for our non-Abelian source.

Most of the time we will write the quark trajectory in Lorentz-covariant notation parametrized by proper time, $x(\tau)\equiv x^{\mu}(\tau)$.
In our analysis below we will have need to refer to
the velocity, acceleration, jerk and snap of the quark, which will be denoted as
\begin{equation}\label{tderivatives}
\vec{v}\equiv {d\vec{x}\over dt}~,\quad
\vec{a}\equiv {d\vec{v}\over dt}~, \quad
\vec{j}\equiv {d\vec{a}\over dt}~, \quad
\vec{s}\equiv {d\vec{j}\over dt}~,
\end{equation}
or, in four-vector form,
\begin{equation}\label{tauderivatives}
v\equiv {dx\over d\tau}~,\quad
a\equiv {dv\over d\tau}~, \quad
j\equiv {da\over d\tau}~, \quad
s\equiv {dj\over d\tau}~,
\end{equation}
which by definition satisfy the relations
\begin{equation}\label{tauidentities}
v^2=-1~,\quad v\cdot a=0~,\quad v\cdot j=-a^2~.
\end{equation}

By the AdS/CFT dictionary, it is the endpoint of the string on the D7-branes that directly corresponds to the quark, whereas the body of the string codifies the profile of the gluonic field sourced by the quark, which is precisely the information we are after. Specifically, our aim is to probe the gluonic field configuration by determining the one-point function of the operator
\begin{equation} \label{o}
\cO_{F^{2}}\equiv
{1\over 2 g_{YM}^{2}}
\tr \left\{
   F_{\mu\nu}F^{\mu\nu}+[\Phi_{I},\Phi_{J}] [\Phi^{I},\Phi^{J}]
  +\mbox{fermions}
\right\}~,
\end{equation}
where $\Phi^{I}$, $I=1,\ldots,6$,
denote the scalar fields of the $\cN=4$ SYM theory. $\cO_{F^{2}}$ is essentially the SYM Lagrangian density. As a shorthand, from this point on we will refer to it (including the normalization constant in front) as simply $\trFsq$. Since, as mentioned above, we work in the approximation where the quark follows a definite trajectory, its presence amounts to the addition of the corresponding Wilson line, due to which we expect the one-point function of $\trFsq$ to be nonvanishing. The latter operator is known to be dual to the IIB (s-wave) dilaton field $\phi(x,z)$ \cite{igor,iwm}, and the standard GKPW recipe for correlation functions \cite{gkpw} at large $N_c$ and $\lambda$ relates its
one-point function to a variation of the supergravity action with respect to the boundary value of $\phi$. The connection can be summarized as \cite{cg}
\begin{equation} \label{trfsq}
\expec{\trFsq(x)}=-
\lim_{z\to 0}\left({1\over z^{3}}\p_{z}\varphi(x,z) \right)~,
\end{equation}
in terms of a rescaled dilaton field $\varphi\equiv\Omega_{5}R^{8}\phi/ 2\kappa^{2}$.

Our task, then, is to calculate the dilaton field sourced by the string
dual to the quark, and to pick out the $O(z^4)$ term in its expansion near the
boundary of AdS$_5$. In the linearized approximation appropriate at large $N_c$, $\varphi$ is obtained simply by convolving the string source with the retarded dilaton propagator \cite{dkk}
\begin{equation} \label{prop}
D(U)={1\over 4\pi^{2}R^{3}}\,{d\over dU} \left[ \frac{2U^2-1}{\sqrt{1-U^2}} \Theta(1-|U|)\right]~,
\end{equation}
which  depends only on the invariant AdS distance between the observation (unprimed) and source (primed) point,
\begin{equation} \label{U}
U \equiv 1- \frac{(t-t')^{2}-(\vx-\vx')^{2}-(z-z')^{2}}{2zz'}
=\frac{(x-x')^2+z^2+z'^2}{2zz'}~.
\end{equation}
The symbol $\Theta$ inside the brackets of (\ref{prop}) denotes the Heaviside step function, which implements causality, as we will elaborate on below.

The string dynamics is prescribed as usual by the Nambu-Goto action
\begin{equation}\label{nambugoto}
S_{\mbox{\scriptsize NG}}=-{1\over 2\pi\ls^2}\int
d^2\sigma\,\sqrt{-\det{g_{ab}}}
\end{equation}
where $g_{ab}\equiv\p_a X^m\p_b X^n G_{mn}(X)$ ($a,b=0,1$) denotes
the induced metric on the worldsheet. We work for the time being in the static gauge
$\sigma^0=t$, $\sigma^1=z$, so the string embedding is described by $\vX(t,z)$. Given that the string endpoint represents the quark, the latter's trajectory is read off from the string profile via
\begin{equation}\label{quarkx}
\vec{x}(t)=\vec{X}(t,z_m)~.
\end{equation}
Since we have mentioned that for $z_m>0$ our fundamental color source is not pointlike, we should be more precise: the AdS/CFT dictionary identifies $\vec{X}(t,z_m)$ as the location of the most UV contribution to the gluonic field, and it is this that we adopt as a natural definition for the position of the quark. This definition has proven to be useful to capture the physics of a range of phenomena including Brownian motion \cite{brownian} and (as we will recall below) radiation damping \cite{lorentzdirac,damping}, but, given the extended nature of the quark, is of course not unique (see, e.g., \cite{jets}).

Using (\ref{prop}), and taking into account the coupling (in the Einstein frame) between the dilaton and the string worldsheet, one finds the relation \cite{cg}
\begin{equation} \label{dilsol}
\varphi(x,z)={1\over 16\pi^{3}\ls^2}\int dt'\,dz'\,\sqrt{-g(t',z')}
  \,{d\over dU}  \left[ \frac{2U^2-1}{\sqrt{1-U^2}} \Theta(1-|U|)\right]~.
\end{equation}
Together with (\ref{trfsq}), this will yield the result we are seeking. To put this to use,
we are missing only one ingredient: we must know the explicit form of the string embedding.

Of course, just like the specification of a quark trajectory does not uniquely determine a gluonic field configuration, knowing the motion of the string endpoint is not enough to select a unique string profile. Additional information is needed, in the form of initial or boundary conditions. For a quark in vacuum, the configuration of most evident interest is the \emph{retarded} one, where waves in the gluonic field move out from the quark to infinity, rather than the other way around, or some (nonlinear) superposition thereof. Luckily, the corresponding solution of the Nambu-Goto equation is known,
for an \emph{arbitrary} time-like quark/endpoint trajectory. In the case where the quark is infinitely massive ($z_m=0$), this solution can be written as \cite{mikhailov}
\begin{eqnarray}\label{mikhsolnoncovariant}
\vec{X}(t_r,z)&=&\vec{x}(t_r)+\frac{\vec{v}(t_r) z}{\sqrt{1-\vec{v}(t_r)^2}}~,\\
t(t_r,z)&=&t_r+\frac{z}{\sqrt{1-\vec{v}(t_r)^2}}~, \nonumber
\end{eqnarray}
 where the behavior of the string at a given time $t$ and radial depth $z$ is seen to be completely determined by the behavior of the quark/string endpoint at the earlier, retarded time $t_r$.\footnote{Reversing the sign of the $z$-dependent terms in (\ref{mikhsol}) yields instead an \emph{advanced} solution \cite{mikhailov}. The form of the most general Nambu-Goto solution for an arbitrary endpoint trajectory is not known.} The definition of $t_r$ implicit in (\ref{mikhsolnoncovariant}) can be shown to follow from projecting back to the AdS boundary along a curve that is null on the string worldsheet \cite{mikhailov}, in analogy with the  Lienard-Wiechert story in classical electrodynamics. In Lorentz-covariant language, the solution (\ref{mikhsolnoncovariant}) takes the simple form
 \begin{equation}\label{mikhsol}
 X^{\mu}(\tau,z)=x^{\mu}(\tau)+zv^{\mu}(\tau)~,
\end{equation}
 where $\tau$ is the \emph{quark} proper time corresponding to $t_r$.\footnote{To avoid confusion, we  emphasize here that $\tau$ is defined with the Minkowski metric appropriate for the SYM theory, and therefore differs from the proper time of the string endpoint (computed with the AdS metric (\ref{metric})) by a factor of $R/z_m$ (where we are currently discussing the case $z_m=0$).}  {}From now on we will mostly work with this parametrization of the string embedding, where the natural notion of proper time $\tau$ associated (modulo a rescaling) with the endpoint has been extended to the full worldsheet by following the upward null geodesics. We should stress that $\tau$ is associated with $t_r$ and is therefore still a \emph{retarded} time parameter, in spite of the fact that, for brevity, we are not labeling it with the  subindex $r$.

For the case of a quark with finite mass (and therefore, finite size), $z_m>0$, the string endpoint is at $z=z_m$, and is subject to the boundary condition (\ref{quarkx}). As in \cite{dragtime,lorentzdirac,damping}, the associated string embeddings  can be regarded as the $z\ge z_m$ portions of the  solutions (\ref{mikhsol}),
which are parametrized by data at the AdS boundary $z=0$. {}From this point on we will use tildes to label
these (now merely auxiliary) data, and distinguish them from the actual physical quantities
(velocity, proper time, etc.) associated with the endpoint/quark at $z=z_m$, which will be denoted
without tildes. In this notation, (\ref{mikhsol}) reads
 \begin{equation}\label{mikhsoltilde}
X^{\mu}(\ttau,z)=\tx^{\mu}(\ttau)+z\tv^{\mu}(\ttau)~.
\end{equation}
As shown in \cite{lorentzdirac,damping},  this can be rewritten purely in
terms of physical ($z=z_m$) data as
\begin{equation}\label{mikhsolzm}
X^{\mu}(\tau,z)=x^{\mu}(\tau)+ \frac{(z-z_m)(v^{\mu}-z^2_m\bar{\cF}^{\mu})}{\sqrt{1-z^4_m \bar{\cF}^2}}~,
\end{equation}
where $\bar{\cF}_{\mu}\equiv (2\pi/\sqrt{\lambda})\cF_{\mu}$, with
$\cF_{\mu}=(-\gamma\vec{F}\cdot\vec{v},\gamma\vec{F})$ the external four-force that needs to be exerted on the quark to get it to follow the prescribed trajectory.\footnote{On the gravity side of the correspondence, this is achieved by subjecting the string endpoint to an electromagnetic field on the D7-branes, and $\cF_{\mu}=-F_{\nu\mu}\p_{\tau}x^{\nu}$ is then the usual Lorentz four-force.} Expression (\ref{mikhsolzm}) clearly reduces to (\ref{mikhsol}) as $z_m\to 0$. As explained in \cite{damping}, the inclusion of the force at finite $z_m$ is a convenient way of summarizing a dependence of the string embedding on an infinite number of higher derivatives of the quark trajectory, whose appearance is natural, given that the dressed quark is an extended object. The AdS/CFT correspondence makes it possible to deduce that the trajectory and the force are connected through the equation of motion
\begin{equation}\label{lorentzdirac}
{d\over
d\tau}\left(\frac{m{d x^{\mu}\over d\tau}-{\sqrt{\lambda}\over 2\pi m}
\cF^{\mu}}{\sqrt{1-{\lambda\over 4\pi^2 m^4}\cF^2}}\right)=\cF^{\mu}-{\sqrt{\lambda}\, \cF^2 \over 2\pi m^2}\left(\frac{{d x^{\mu}\over d\tau}-{\sqrt{\lambda}\over
2\pi m^2} \cF^{\mu} }{1-{\lambda\over 4\pi^2 m^4}\cF^2}\right) ~,
\end{equation}
which incorporates the effects of radiation damping and is thus a nonlinear generalization of the classic Lorentz-Dirac equation, codifying (inside the parentheses on the left-hand side) a non-standard dispersion relation for the dressed quark, as well as (in the second term on the right-hand side) a Lorentz-covariant formula for its rate of radiation \cite{lorentzdirac,damping}.

In the following section, we will use the string embedding (\ref{mikhsolzm}) in (\ref{dilsol}) to determine the resulting dilaton profile, and then extract from (\ref{trfsq}) the desired expectation value of the gluonic field generated by the quark.

\section{Gluonic Profile for Arbitrary Quark Motion}\label{arbitrarysec}

We now proceed to the calculation. If we employ the more physical parametrization $(\tau,z)$ of the string worldsheet, with $\tau$ the (retarded) quark proper time, the string profile takes the form (\ref{mikhsolzm}).  Computing the induced metric on the worldsheet, one finds that
\begin{equation}\label{g}
\sqrt{-g(\tau',z')}=\frac{R^2}{z'^2\sqrt{1-z^4_m \bar{\cF}^2(\tau')}}~.
\end{equation}
We have found it more convenient to carry out our entire computation with the worldsheet parametrized by $(\ttau,z)$, with $\ttau$ the proper time of the auxiliary (and fictitious) string endpoint at $z=0$, under which the embedding takes the simpler form (\ref{mikhsoltilde}), implying
\begin{equation}\label{gtilde}
\sqrt{-g(\ttau',z')}=\frac{R^2}{z'^2}
\end{equation}
(which naturally coincides with the $z_m\to 0$ limit of (\ref{g})), and similarly, all subsequent expressions turn out to be more compact. Once we obtain a final result for the expectation value of $\trFsq$, we will show how to eliminate all auxiliary (tilde) variables in favor of the physical (nontilde) ones.

{}Given (\ref{gtilde}) and (\ref{lambda}), to determine the dilaton field (\ref{dilsol}) we need to compute the double integral
\begin{equation} \label{dilsolmikh}
\varphi(x,z)={\sqrt{\lambda}\over 16\pi^{3}}\int_{-\infty}^{\infty} d\ttau'\int_{z_m}^{\infty}{dz'\over z'^2}
  \,{d\over dU}  \left[ \frac{2U^2-1}{\sqrt{1-U^2}} \Theta(1-|U|)\right]~,
\end{equation}
where the invariant distance (\ref{U}) with the source point taken on the embedding (\ref{mikhsoltilde}) adopts the form
\begin{equation} \label{Umikh}
U =\frac{\left(x-X(\ttau',z')\right)^2+z^2+z'^2}{2zz'}
=\frac{(x-\tx(\ttau'))^2+z^2}{2zz'}-\frac{(x-\tx(\ttau'))\cdot\tv(\ttau')}{z}~.
\end{equation}
Notice how the $z'^2$ term of the numerator, present in the definition (\ref{U}), has exactly canceled due to the $z'$-dependence of the retarded solution (\ref{mikhsoltilde}). This cancelation will play a crucial role in the form of our results, and we will return to it in Section \ref{discussionsec}.

The presence of the $U$-derivative in (\ref{dilsolmikh}) suggests a change of variables $z'\to U$. {}From (\ref{Umikh}) we see that
\begin{equation}\label{dU}
dU=-dz'\left(\frac{(x-\tx(\ttau'))^2+z^2}{2zz'^2}\right)~,
\end{equation}
so we are left with\footnote{Upon carrying out the $U$ integral to be left with the surface term, one might worry about the fact that the quotient in front of the Heaviside function diverges at the endpoints of the interval that the latter defines. This concern can be dispelled by noting that $\Theta(1-|U|)=\Theta(1-U)\Theta(1+U)$ appears \emph{within} the $U$ derivative, and upon differentiation via the Leibniz rule, would give rise to delta functions that precisely cancel these divergences.}
\begin{eqnarray} \label{dilsolmikh2}
\varphi(x,z)&=&-{\sqrt{\lambda}\over 8\pi^{3}}\int_{-\infty}^{\infty}
{ d\ttau'\, z\over (x-\tx(\ttau'))^2+z^2}
\int_{\Umin}^{\Umax}\!\!dU
  \,{d\over dU}  \left[ \frac{2U^2-1}{\sqrt{1-U^2}} \Theta(1-|U|)\right]\nonumber\\
&=&  -{\sqrt{\lambda}\over 8\pi^{3}}\int_{-\infty}^{\infty}
{ d\ttau'\, z\over (x-x(\ttau'))^2+z^2}
 \left[ \frac{2U^2-1}{\sqrt{1-U^2}} \Theta(1-|U|)\right]_{\Umin}^{\Umax}~,
\end{eqnarray}
where we have defined
\begin{eqnarray}\label{Uminmax}
\Umin&\equiv&\frac{(x-\tx(\ttau'))^2+z^2}{2zz_m}-\frac{(x-\tx(\ttau'))\cdot\tv(\ttau')}{z}~,\\
\Umax&\equiv&-\frac{(x-\tx(\ttau'))\cdot\tv(\ttau')}{z}~.\nonumber
\end{eqnarray}

\begin{figure}[htb]
\begin{center}
\setlength{\unitlength}{1cm}
\includegraphics[width=12cm,height=6cm]{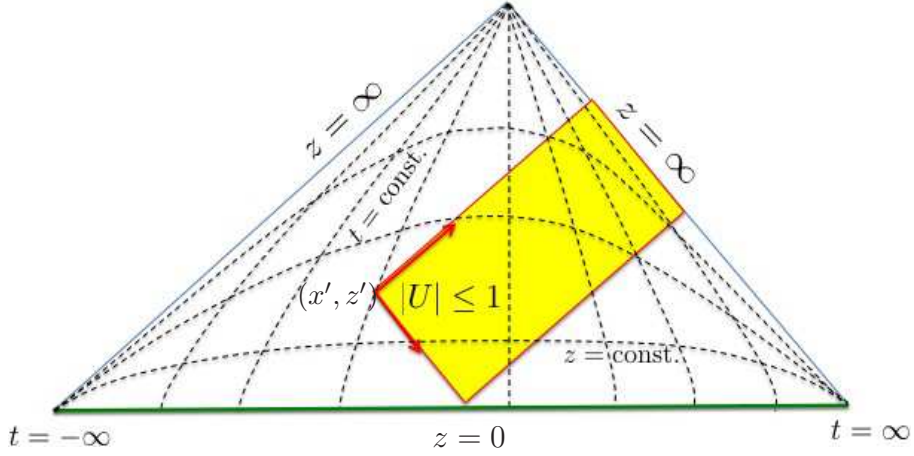}
 \begin{picture}(0,0)
 \put(-8.3,1.9){\small $(x',z')$}
 \put(-6.5,0){$z=0$}
\end{picture}
\caption{Penrose diagram of AdS (with boundary directions other than the time omitted), showing a Poincar\'e coordinate grid. The shaded yellow region $|U|\le 1$ indicates the observation points $(x,z)$ that can be influenced by the given source point $(x',z')$.}
\label{penrosefig}
\end{center}
\end{figure}

It is now necessary to understand the physical import of the $\Theta(1-|U|)$ factor. Its presence in (\ref{prop}) implies that the propagator $D(U)$ has the geometric structure shown in Fig.~\ref{penrosefig}: a given source point $(x',z')$ induces a nonvanishing dilaton on and inside its forward lightcone, up to the event where this lightcone reflects back from the boundary. Beyond that, there are some observation points $(x,z)$ that lie in the future of the source point but are not influenced by it. The region $U>1$ corresponds to observation points that are in the causal past of the event $(x',z')$ or are spacelike related to it. Causality dictates that these points will be beyond the region of influence of the source.  Exclusion of points at $U<-1$, on the other hand, is \emph{a priori} unexpected, because they do obey the naive causality restriction, and  are only disallowed due to conditions at the boundary of AdS. These are points that can be reached by a timelike curve originating at $(x',z')$, but not by a timelike \emph{geodesic}.

\begin{figure}[htb]
\begin{center}
\setlength{\unitlength}{1cm}
\includegraphics[width=12cm,height=6cm]{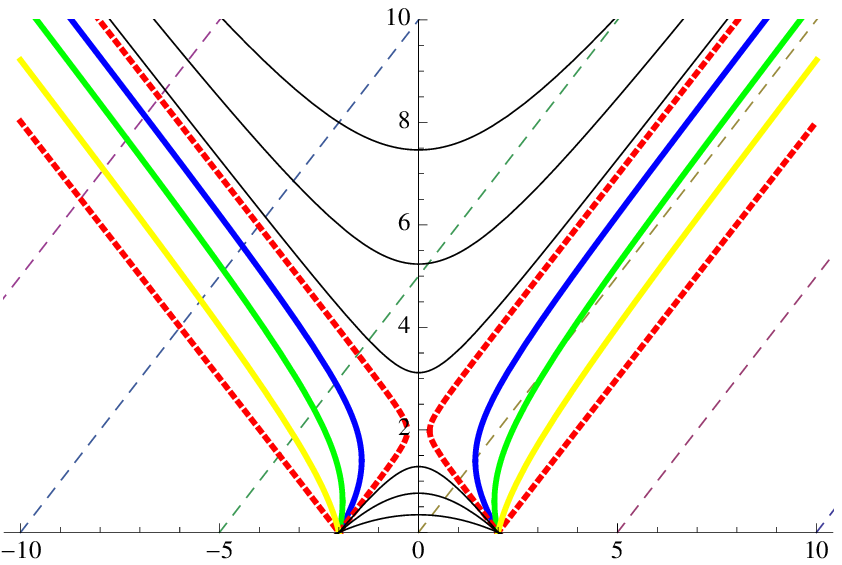}
 \begin{picture}(0,0)
 \put(-6,6.1){\small $z'$}
   \put(-2.2,2.7){\small $U=-1$}
    \put(-6.7,1.7){\small $U=1$}
       \put(-7.6,3.9){\small $U=1.5$}
       \put(-8.3,5.4){\small $U=2$}
   \put(-12.7,5.7){\small $U=0$}
  \put(-11.4,2.7){\small $U=-1$}
   \put(-2.5,0.7){\small $\ttau'=5$}
  \put(-10.1,0.7){\small $\ttau'=-5$}
     \put(-12.5,1.3){\small $\ttau'=-10$}
   \put(-13.2, 3.3){\small $\ttau'=-15$}
 \put(0,0){$t'$}
\end{picture}
\caption{For a static quark, this is the way in which our new coordinates $(\ttau',U)$ cover the string worldsheet which we had originally parametrized with $(t',z')$. A detailed explanation is given in the paragraph before (\ref{dilsolmikh3}). For ease of visualization, we have chosen the observation point at $x^{\mu}=0$, $z=2$ (even though we are ultimately interested in $z\to 0$). The causal swath $|U|\le 1$ is the region between the 2 dotted (red) hyperbolae on the left (which collapses to a line as $z\to 0$).}
\label{hyperbolasfig}
\end{center}
\end{figure}

In Fig.~\ref{hyperbolasfig} we show the way in which our new coordinates $(\ttau',U)$ cover the string worldsheet, which we had originally parametrized with $(t',z')$. For concreteness, the coordinate grid has been plotted for the case where the quark is static; for more general motions the overall pattern is similar but somewhat distorted. As mentioned above, the constant $\ttau'$ (or equivalently, constant $t'_r$) curves are null; on the $(t',z')$ plane they are simply straight lines with slope $1/\gamma$. The range of constant-$U$ curves that intersect a given constant-$\ttau'$ line runs from $U=\Umin(\ttau';x,z)$, where the intersection takes place at $t'=t'_r$, $z'=0$, to $U=\Umax(\ttau';x,z)$, which is asymptotic to the line and only intersects it at $t',z'\to\infty$.  {}From this it becomes clear that, for any finite $\ttau'$ and any finite observation point, the point $U=\Umax$ lies outside of the causal swath $-1\le U\le 1$ allowed by the Heaviside function. Consequently, only the $U=\Umin$ endpoint contributes to (\ref{dilsolmikh2}), leaving us with
\begin{equation} \label{dilsolmikh3}
\varphi(x,z)=  {\sqrt{\lambda}\over 8\pi^{3}}\int_{-\infty}^{\infty}
{ d\ttau'\, z\over (x-\tx(\ttau'))^2+z^2}
 \left[ \frac{2\Uminsq-1}{\sqrt{1-\Uminsq}} \Theta(1-|\Umin|)\right]~.
\end{equation}

To carry out the remaining integral, it is convenient to perform a final change of variables $\ttau'\to\Umin$. {}From (\ref{Uminmax}) we can deduce, via implicit differentiation, that
\begin{equation}\label{dUmin}
d\ttau'=-\frac{z\, z_m\, d\Umin}{z_m+(x-\tx(\ttau'))\cdot(\tv(\ttau')+ z_m\ta(\ttau'))}~,
\end{equation}
but proceeding further requires explicitly inverting (\ref{Uminmax}), which is impossible to do exactly for an arbitrary trajectory. We must remember from (\ref{trfsq}), however, that we are ultimately only interested in the value of $\varphi(x,z)$ in the limit $z\to 0$, up to order $z^4$. In this limit the causal swath $-1\le U\le 1$ depicted in Fig.~\ref{hyperbolasfig} in fact collapses to a curve on the $(t',z')$ plane, but the function inside the brackets in (\ref{dilsolmikh3}) still varies wildly over the allowed $\Umin$ range, making it necessary to really carry out the integral instead of just evaluating the integrand at any point within the interval and multiplying times the interval width. A natural strategy is then to Taylor-expand all expressions in powers of $z$. Using (\ref{dUmin}) in (\ref{dilsolmikh3}), we see that we already have a factor of $z^2$ up front, so we only need to carry out the expansions to order $z^2$, because terms of order $z^3$ and higher will not contribute to the one-point function in the $z\to 0$ limit. We start by proposing that
\begin{equation}\label{ttautaylor}
\ttau'=\ttau_0+\ttau_1 z + \ttau_2 z^2 + \cO(z^3)~,
\end{equation}
where $\ttau_j=\ttau_j(x,\Umin)$ are independent of $z$. Using (\ref{ttautaylor}) we can deduce that
\begin{equation}\label{xtaylor}
\tx(\ttau')=\tx(\ttau_0)+\ttau_1 \tv(\ttau_0)z + \left(\ttau_2 \tv(\ttau_0) + {1\over 2}\ttau_1^2\ta(\ttau_0)\right)z^2 + \cO(z^3)~,
\end{equation}
and similarly for $\tv(\ttau')$. With this information in hand, we can Taylor-expand (\ref{Uminmax}) in a power series in $z$, match terms of the same order and solve recursively to find the coefficients $\ttau_j$. At leading order we get a condition determining $\ttau_0$,
\begin{equation}\label{ttau0}
(x-\tx(\ttau_0))^2=2 z_m (x-\tx(\ttau_0))\cdot\tv(\ttau_0)~.
\end{equation}
Proceeding to higher order one finds subsequently
\begin{eqnarray}\label{ttaucoeffs}
\ttau_1&=&-\frac{\Umin \, z_m}{z_m+(x-\tx')\cdot(\tv'+\ta' z_m)}~,\\
\ttau_2&=&\frac{1}{2\left[z_m+(x-\tx')\cdot(\tv'+\ta' z_m)\right]}
-\frac{\Uminsq z_m^2\left(1+(x-\tx')\cdot(\ta'+\tj' z_m)\right)}{2\left[z_m+(x-\tx')\cdot(\tv'+\ta' z_m)\right]^3}~, \nonumber
\end{eqnarray}
where from now on it is understood that functions denoted with a prime are evaluated at the retarded time $\ttau=\ttau_0$ defined implicitly by (\ref{ttau0}).

Using (\ref{dUmin})-(\ref{ttaucoeffs}) in (\ref{dilsolmikh3}), one can Taylor expand in powers of $z$, and carry out the integral over $\Umin$ to determine $\varphi(x,z)$. The master formula (\ref{trfsq}) then leads to the gluonic profile we were after,
\begin{eqnarray}\label{trfsqfinal}
\expec{\trFsq(x)}&=&{\sqrt{\lambda}\over 32\pi^{2}}
\frac{1}{\left((x-\tx')\cdot\tv'\right)^2 \left[z_m+(x-\tx')\cdot(\tv'+\ta' z_m)\right]^5}\\
{}&{}&\quad \times\left\{2\left((x-\tx')\cdot\tv'\right)^3
             +z_m^3 \left(1+(x-\tx')\cdot\ta'\right)^3\right.\nonumber\\
{}&{}&\quad\quad\; +z_m\left((x-\tx')\cdot\tv'\right)^2\left[2+(x-\tx')\cdot(2\ta'-4\tj' z_m-\ts' z_m^2)-\ta'^2 z_m^2\right]\nonumber\\
{}&{}&\quad\quad\; +z_m^2(x-\tx')\cdot\tv'\left[4+4\left((x-\tx')\cdot\ta'\right)^2
+2(x-\tx')\cdot\tj'z_m\right.\nonumber\\
{}&{}&\quad\quad\qquad\qquad\quad
    +3\left((x-\tx')\cdot\tj'\right)^2 z_m^2 -(x-\tx')\cdot\ts'z_m^2-\ta'^2 z_m^2
\nonumber\\
{}&{}&\quad\quad\qquad\qquad\quad
   +
   (x-\tx')\cdot\ta'\left(8+(x-\tx')\cdot(2\tj'-\ts'z_m)z_m-\ta'^2 z_m^2\right)\Big]\bigg\}~.
   \nonumber
\end{eqnarray}
The calculation is straightforward but a bit messy, so we defer the details to the Appendix.

Now that we have our desired result, all that remains is to rewrite the auxiliary variables $\tx$, $\tv$, $\ta$, $\tj$ and $\ts$, which describe the motion of a fictitious endpoint at $z=0$, in
terms of the real physical variables $x$, $v$, $a$, $j$ and $s$, associated with the quark/endpoint at $z=z_m$. This connection has been worked out in \cite{damping}. Equation (25) of that paper states that $\ta$ translates into nontilde variables according to
\begin{equation}\label{atilde}
\ta^{\mu}=\frac{z_m}{\sqrt{1-z^4_m \bar{\cF}^2}}\left(\bar{\cF}^{\mu}-z^2_m \bar{\cF}^2v^{\mu}\right)~,
\end{equation}
where  $\bar{\cF}^{\mu}$ is the rescaled version of  the external four-force applied on the quark that we introduced below (\ref{mikhsolzm}). {}From this and equation (20) of \cite{damping} it also follows that
\begin{equation}\label{vtilde}
\tv^{\mu}=\frac{v^{\mu}-z^2_m\bar{\cF}^{\mu}}{\sqrt{1-z^4_m \bar{\cF}^2} }~,
\end{equation}
which in turn implies, via (\ref{mikhsoltilde}),
\begin{equation}\label{xtilde}
\tx^{\mu}=x^{\mu}-\frac{z_m(v^{\mu}-z^2_m\bar{\cF}^{\mu})}
{\sqrt{1-z^4_m \bar{\cF}^2} }~.
\end{equation}
Comparing (\ref{g}) and (\ref{gtilde}) we see that
\begin{equation}\label{tau2}
 d\ttau=\frac{d\tau}{\sqrt{1-z^4_m \bar{\cF}^2}}~,
 \end{equation}
which enables us to deduce from (\ref{atilde}) that
\begin{multline}\label{jtilde}
\tj^{\mu}= z_m\,\dot{\bar{\!\cF}}^{\mu}-z^3_m\bar{\cF}^2a^{\mu}
+\frac{z_m^5(\bar{\cF} \cdot \,\dot{\bar{\!\cF}})\bar{\cF}^{\mu}}{(1-z^4_m \bar{\cF}^2)}
-\Big{[}z^3_m(\bar{\cF} \cdot \,\dot{\bar{\!\cF}})+\frac{z^7_m(\bar{\cF} \cdot \,\dot{\bar{\!\cF}}){\bar{\cF}}^2}{(1-z^4_m \bar{\cF}^2)}\Big{]}v^{\mu}
\end{multline}
and
\begin{multline}\label{stilde}
\ts^{\mu}=\Big{[}z_m\sqrt{1-z^4_m \bar{\cF}^2}
\,\ddot{\bar{\!\cF}}^{\mu}+\frac{z^5_m}{(1-z^4_m \bar{\cF}^2)^{3/2}}\Big{(((\bar{\cF} \cdot \,\ddot{\bar{\!\cF}})+{\,\dot{\bar{\!\cF}}}^2})(1-z^4_m \bar{\cF}^2)\bar{\cF}^{\mu}\\
(\bar{\cF} \cdot \,\dot{\bar{\!\cF}})(1-z^4_m \bar{\cF}^2)\,\dot{\bar{\!\cF}}^{\mu}+2z^4_m(\bar{\cF} \cdot \,\dot{\bar{\!\cF}})^2\bar{\cF}^{\mu}\Big{)}\Big{]}\\
-\frac{z^3_m}{(1-z^4_m \bar{\cF}^2)^{3/2}}\Big{[(\bar{\cF} \cdot \,\ddot{\bar{\!\cF}})(1-z^4_m \bar{\cF}^2)+ \,\dot{\bar{\!\cF}}^2(1-z^4_m \bar{\cF}^2)+2z^4_m(\bar{\cF} \cdot \,\dot{\bar{\!\cF}})^2}\Big{]}v^{\mu}\\
-\Big{[}\frac{z^3_m(\bar{\cF} \cdot \,\dot{\bar{\!\cF}})}{\sqrt{1-z^4_m \bar{\cF}^2}}+2z^3_m\sqrt{1-z^4_m \bar{\cF}^2}(\bar{\cF} \cdot \,\dot{\bar{\!\cF}})\Big{]}a^{\mu}
-z^3_m \bar{\cF}^2\sqrt{1-z^4_m \bar{\cF}^2} \bar{\cF}^2j^{\mu}~,
\end{multline}
where dots over the four-force denote $\tau$ derivatives.

Employing (\ref{atilde}), (\ref{vtilde}), (\ref{xtilde}), (\ref{jtilde}) and (\ref{stilde}), we can express the profile (\ref{trfsqfinal}) of the gluonic field directly in terms of the physical (nontilde) variables that describe the quark trajectory. This is our main result. We refrain from writing the resulting expression, because it is long and not particularly enlightening.

All dynamical quantities in the final result are understood to be evaluated at the retarded proper time $\tau_0$ defined by translating (\ref{ttau0}) into physical variables, i.e.,
\begin{equation}\label{tau0}
(x-x(\tau_0))^2=-z_m^2~.
\end{equation}
This equation describes a two-sheeted hyperboloid about the observation point $x$, which is intersected by the quark worldline twice, once on each sheet. By causality, the root of interest is of course the one in the sheet to the past of $x$, which, in noncovariant notation, corresponds to the retarded time
\begin{equation}\label{tret}
\tret=t-\sqrt{(\vec{x}-\vec{x}(\tret))^2+z_m^2}~.
\end{equation}
This past hyperboloid describes all the points $x'=(\tret,\vec{x}(\tret))$ that can influence the given observation point. We can of course read (\ref{tau0}) the other way around, as a statement that the events $x$ that can be influenced by a given source point $x'=x(\tau_0)$ on the quark worldline, lie on the future hyperboloid at constant timelike interval $z_m$ from $x'$.

Before moving on to applications and a discussion on the physical content of our result, let us perform a check on it, by examining what it implies for a free quark. In the absence of external forcing, the equation of motion (\ref{lorentzdirac}) naturally implies that the quark moves at constant velocity (although, interestingly, the converse is not true \cite{damping}). {}From (\ref{atilde})-(\ref{stilde}), we have $\tx=x-z_m v$, $\tv=v$, $\ta=\tj=\ts=0$, and therefore (\ref{trfsqfinal}) reduces to
\begin{equation}\label{trfsqconstantv}
\expec{\trFsq(x)}={\sqrt{\lambda}\over 32\pi^{2}}
\frac{2\left[(x-x')\cdot v'\right]^3-4\left[(x-x')\cdot v'\right]^2 z_m
+6\left[(x-x')\cdot v'\right]z_m^2-3z_m^3}
{\left((x-x')\cdot v'-z_m\right)^2\left[(x-x')\cdot v'\right]^5}
\end{equation}
By Poincar\'e invariance, it suffices to evaluate this in the frame where the quark is at rest at the origin. Using (\ref{tret}), we then have $(x-x')\cdot v'=-(t-\tret)=-\sqrt{\vec{x}\,^2+z_m^2}$, so we are left with
\begin{equation}\label{trfsqstatic}
\expec{\trFsq(x)}={\sqrt{\lambda}\over 32\pi^{2}}
\frac{4\vec{x}\,^2 z_m+7z_m^3
+(2\vec{x}\,^2+8z_m^2)\sqrt{\vec{x}\,^2+z_m^2}}
{\left(\sqrt{\vec{x}\,^2+z_m^2}+z_m\right)^2\left[\vec{x}\,^2+z_m^2\right]^{5/2}}~.
\end{equation}
This can be further massaged into the form
\begin{equation}\label{trfsqstatic2}
\expec{\trFsq(x)}={\sqrt{\lambda}\over 16\pi^{2}\vec{x}\,^4}
\left(1-\frac{z_m^3\left(z_m^2+{5\over 2}\vec{x}\,^2\right)}{\left[\vec{x}\,^2+z_m^2\right]^{5/2}}\right)~,
\end{equation}
which precisely coincides with the gluonic profile determined in \cite{martinfsq} for a static quark with finite mass.

For future use we also note that in the limit $z_m\to 0$, where the quark becomes pointlike (as well as infinitely massive), our general result (\ref{trfsqfinal}) simplifies drastically, and we are left with the single term
\begin{equation}\label{trfsqpointlike}
\expec{\trFsq(x)}={\sqrt{\lambda}\over 16\pi^{2}}
\frac{1}{\left[(x-x(\tau_0))\cdot v(\tau_0)\right]^4}~,
\end{equation}
which according to (\ref{tau0}) is to be evaluated at $\tau_0$ such that $(x-x(\tau_0))^2=0$. In other words, in this case signals propagate purely along null intervals: the hyperboloid we had for $z_m>0$ converges to the lightcone, and the retarded time (\ref{tret}) is now at the point of intersection between the quark worldline and the past lightcone of $x$, i.e., $\tret=t-|\vec{x}-\vec{x}(\tret)|$.

\section{Some Applications}\label{examplessec}

\subsection{Harmonic motion} \label{oscillatesubsec}

As a first application of our general result (\ref{trfsqfinal}), we will study the example of a quark undergoing one-dimensional harmonic motion. This is the setup where, under a linearized approximation to the string dynamics, the authors of \cite{cg} obtained a propagation pattern with a broad tail, seemingly in conflict with the no-broadening result of \cite{iancu1,iancu2} and the present paper, but, on the other hand, in consonance with general expectations for a strongly-coupled non-Abelian system \cite{cg,iancu1,iancu2}. In what follows, we will show in particular that, despite appearances, the result reported in \cite{cg} correctly reproduces the linearized limit of our own result.

Let $x\equiv x^1$ denote  the  direction of oscillation. The trajectory is then given by
\begin{equation}\label{xosc}
x^{\mu}=(t, A \sin(\omega t),0,0)~,
\end{equation}
with corresponding four-velocity
\begin{equation}\label{vosc}
v^{\mu}=(\gamma, \gamma \omega A \cos(\omega t),0,0)~,\quad
\gamma=\frac{1}{\sqrt{1-\omega^2A^2\cos^2(\omega t)}}~.
\end{equation}

We will consider first the case (not examined by \cite{cg,iancu1,iancu2}) where the quark has a finite mass. According to (\ref{trfsqfinal}), determining the gluonic profile in this case requires, beyond the position, velocity, acceleration and jerk of the quark, knowledge of the total external force applied to it, together with its first and second derivatives, sometimes called  yank and  tug. The generalized Lorentz-Dirac equation (\ref{lorentzdirac}) specifies the four-force $\cF^{\mu}\equiv\gamma(\vec{F}\cdot\vec{v},\vec{F})$ that corresponds to any given quark trajectory. For one-dimensional harmonic motion, this equation demands that $F\equiv F^1$ satisfy
\begin{equation}\label{lorentzdiracoscillate}
\frac{d\bar{F}}{dt}=-\sqrt{1-\omega^2A^2\cos^2(\omega t)}\sqrt{1-z_m^4\bar{F}^2}{\bar{F}\over z_m}
-\frac{\omega^2 A \sin(\omega t)}{1-\omega^2A^2\cos^2(\omega t)}
\left(\frac{1-z_m^4\bar{F}^2}{z_m^2}\right).
\end{equation}
Notice that the nonlinearity of this equation implies that the force is in general \emph{not} harmonic, as a consequence both of the extended nature of the quark and of the damping effect due to the emitted radiation.

In the textbook analysis of the forced and damped harmonic oscillator, the forcing is prescribed to be harmonic and the oscillator motion is synchronized with this forcing only after the decay of a transient component. Conversely, purely harmonic motion would be associated with a force that is initially not harmonic. Numerical exploration of (\ref{lorentzdiracoscillate}) reveals similar behavior: if we prescribe the motion of the quark to be purely harmonic, as in (\ref{xosc}), then for generic initial condition, the force contains a transient component that dies down in a time interval of order $z_m$. At late times, the force is found to be purely oscillatory, but not quite harmonic. Now, as shown in \cite{lorentzdirac,damping} and recalled in Section~\ref{recipesec}, the dependence of the string embedding on the external force codifies an infinite number of higher derivatives of the quark's trajectory, so the initial condition supplied to the numerical integration of (\ref{lorentzdiracoscillate}) at any finite time should in fact be deduced from the behavior of the quark at previous times. If we truly want to explore the situation where the quark has been undergoing harmonic motion \emph{at all times}, then we are automatically forced to work in the late-time force regime where there is no transient. Equivalently, we must fine-tune the initial condition so that the force is purely oscillatory from the start.

To use equation (\ref{trfsqfinal}), we also need to solve (\ref{tau0}) or its noncovariant equivalent (\ref{tret}) to obtain the retarded time $\tau_0$ or $\tret$ at which all dynamical data are to be evaluated. For harmonic quark motion with arbitrary oscillation amplitude, this equation is transcendental, and must also be solved numerically.  Using the information from both numerical integrations in our general expression (\ref{trfsqfinal}), we have obtained the gluonic field profile shown in Fig.~4, which shows that the oscillatory motion generates a non-linear wave, with crests splitting off from the gluonic cloud of the quark every half-cycle. As stipulated by (\ref{trfsqfinal}), the overall pattern decays very fast ($\propto 1/|\vec{x}|^4$), showing no sign of the characteristic radiation falloff.
The waves seen in Fig.~4 are thus fluctuations in the near-field of the quark.

\begin{figure}[htb]
$$
\begin{array}{cc}
  \epsfig{width=2.5in,file=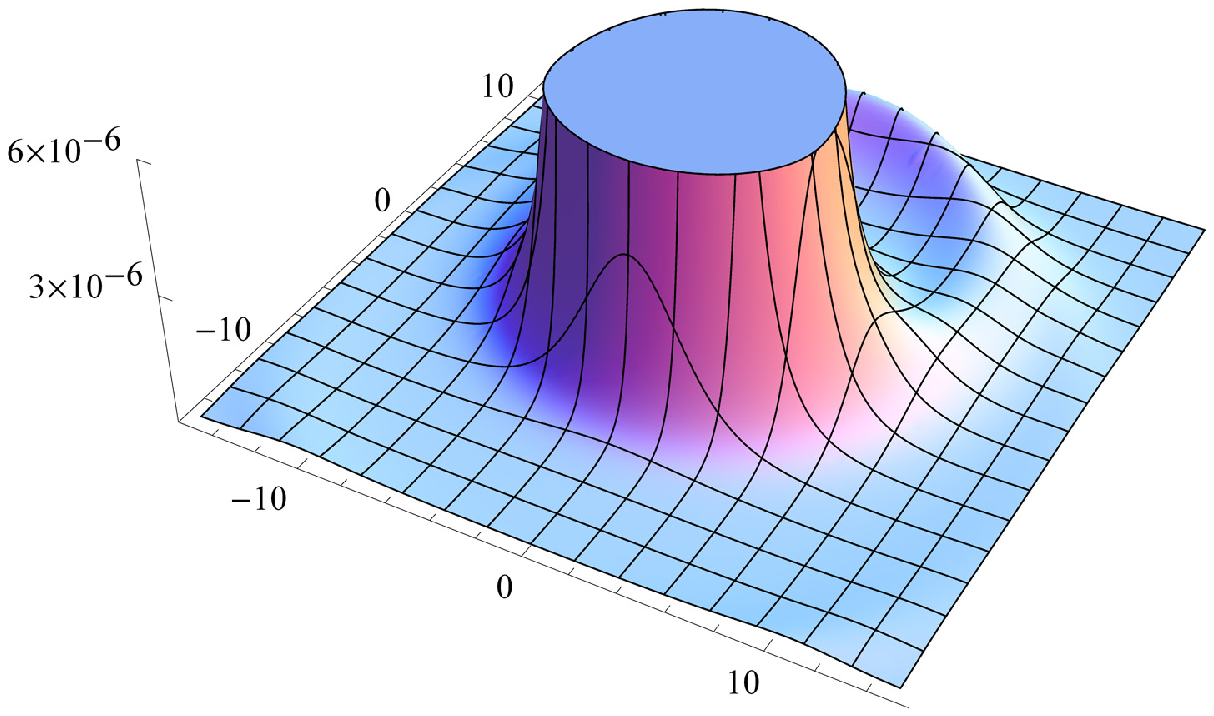} & \epsfig{width=2.5in,file=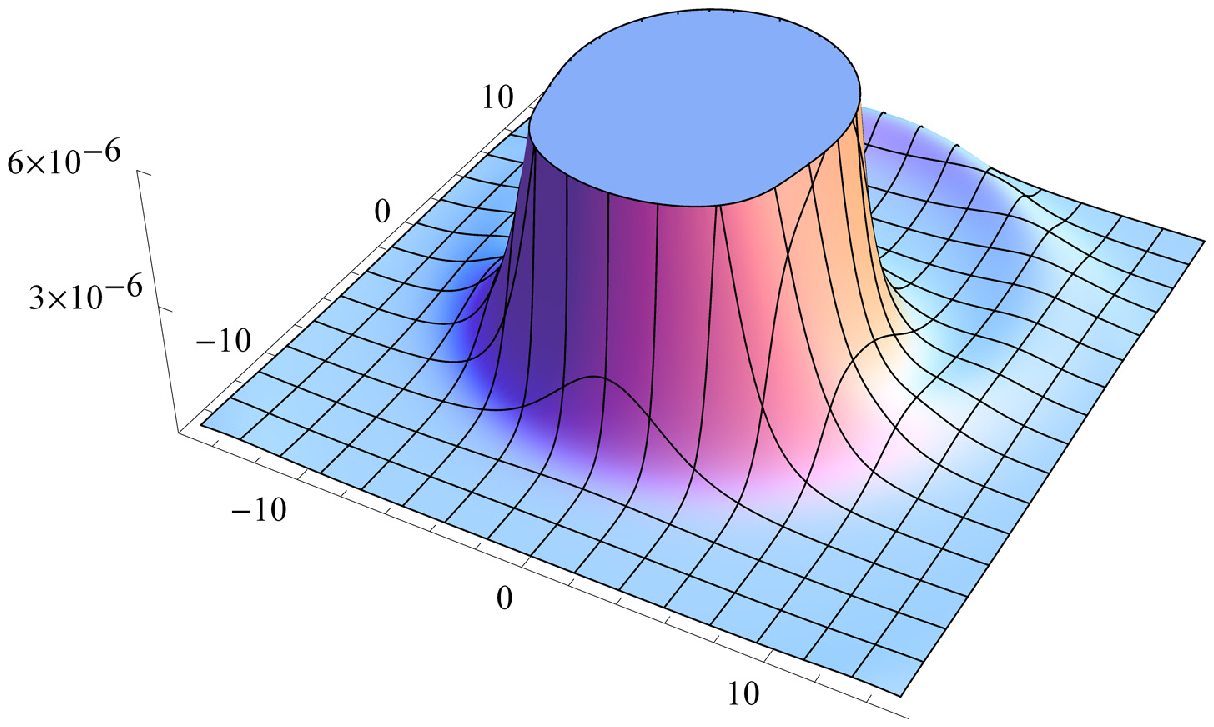} \\
  \epsfig{width=2.5in,file=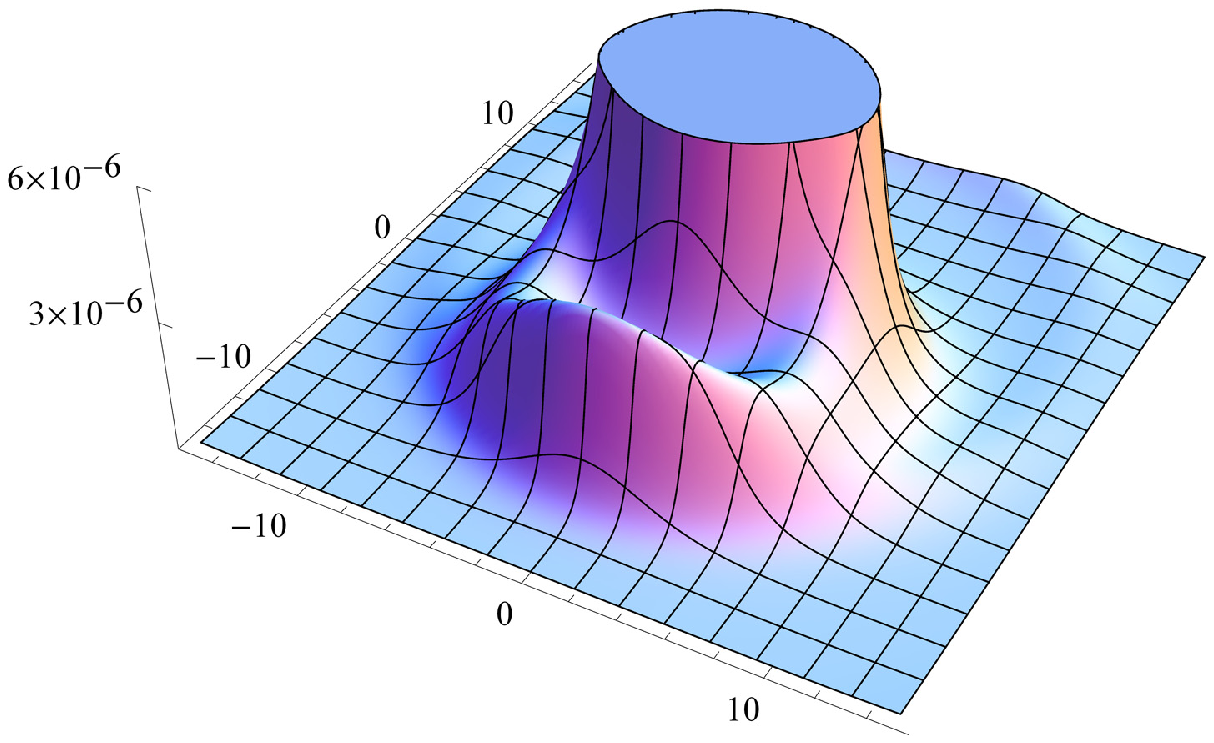} & \epsfig{width=2.5in,file=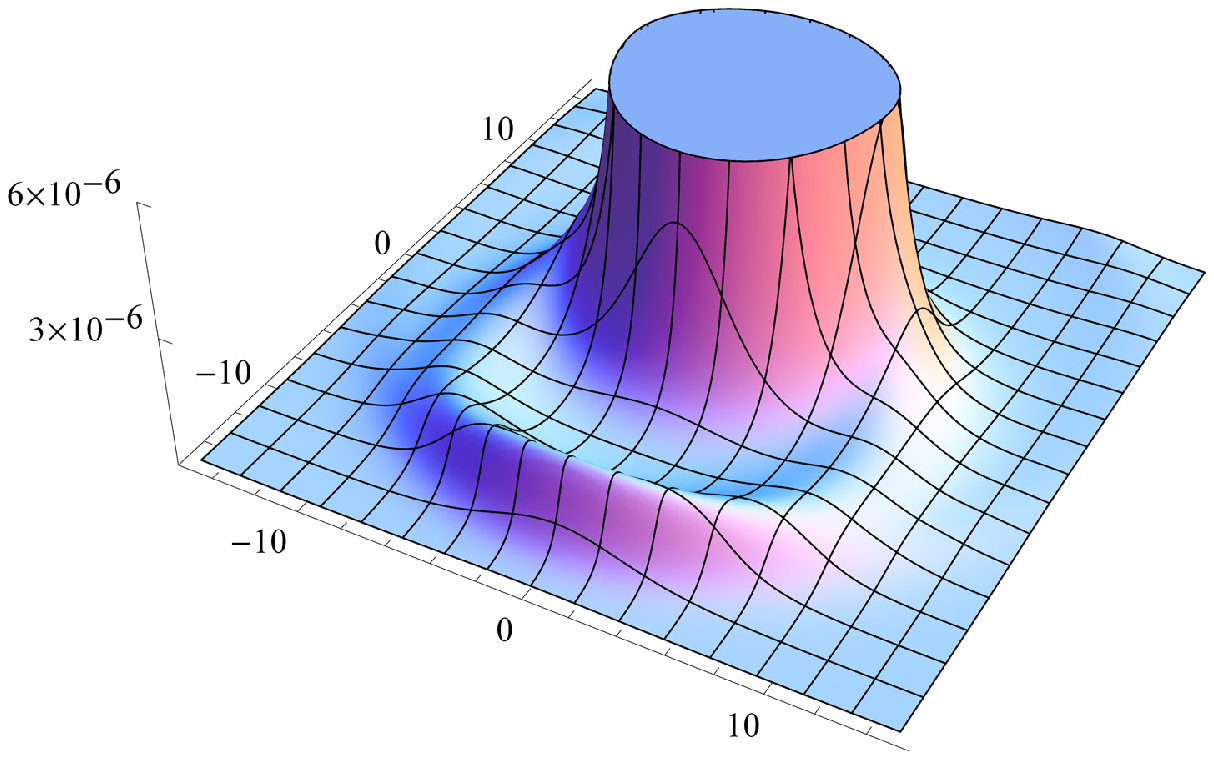}
\end{array}
$$
\begin{picture}(0,0)
 \put(14,78){\tiny $\expec{\tr F^2}$}
 \put(81,78){\tiny $\expec{\tr F^2}$}
 \put(14,36.5){\tiny $\expec{\tr F^2}$}
 \put(81,36.5){\tiny $\expec{\tr F^2}$}
 \put(94,73.5){\tiny $x$}
 \put(27,73.5){\tiny $x$}
 \put(94,32){\tiny $x$}
 \put(27,32){\tiny $x$}
 \put(101.5,49.5){\tiny $y$}
 \put(34.5,49.5){\tiny $y$}
 \put(101.5,8){\tiny $y$}
 \put(34.5,8){\tiny $y$}
 \end{picture}
\vspace{-0.8cm}
\caption{\small Gluonic profile of a heavy quark undergoing harmonic motion in the $x$-direction, with the horizontal axes in units of the quark Compton wavelength, $z_m=1$, and the vertical axis in units of $\sqrt{\lambda}$. We choose sample values  $A=1$ (of the same order as the quark size) and $\omega=1/2$. The plots correspond to four successive snapshots of our simulation, for $t=0,2,4$ and $6$, respectively (all within one oscillation period).  The finite but proportionately large near-field of the quark is capped off in order for the vertical axis to remain at a scale where a non-linear wave pattern is clearly visible, with crests splitting off from the gluonic cloud of the quark every half-cycle.}
\end{figure}

For infinite mass, the quark becomes pointlike and radiation damping becomes negligible. The equation of motion (\ref{lorentzdiracoscillate}) then simplifies drastically, and implies that (up to an additive integration constant)
\begin{equation*}
F(t)={\sqrt{\lambda}\over 2\pi z_m^2}\mbox{arctanh}[A\omega\cos(\omega t)]~,
\end{equation*}
i.e., the required external force is purely oscillatory, but (for arbitrary quark oscillation amplitude) still not harmonic. Nonetheless, we know that for a pointlike quark our formula for the gluonic profile reduces to (\ref{trfsqpointlike}), which depends on the assumed harmonic motion of the quark, but is independent of the external force.

As promised in the introduction, we will now show how to make contact between our results an those of \cite{cg}, which likewise determined the expectation value of $\trFsq$ for an infinitely massive quark undergoing harmonic motion. The authors of that paper carried out their calculation directly in the static gauge $\sigma^0=t$, $\sigma^1=z$, but of course, the fact that we use a different (more geometric) choice of coordinates here cannot affect the final result. A difference that does matter is that \cite{cg} considered only the case where the quark oscillation amplitude is small, $ A\ll 1/\omega$, and thereby treated the string dynamics in a linearized approximation around the static vertical embedding. By inspection of the solution given in \cite{cg}, or equivalently, from the full solution (\ref{mikhsol}), we see that the fluctuation on the string (together with its derivatives) remains small  only for $z\ll 1/\omega $, which thus stipulates the region of validity of the linearized approach. Through the UV/IR connection, this implies that the result of \cite{cg} for the gluonic profile is quantitatively reliable only for observation points that are close to the source, in the sense that $|\vec{x}|\ll 1/\omega $. Altogether then, we will want to examine our general SYM result in the small-amplitude and small-distance (or, equivalently, low-frequency) regime
\begin{equation}\label{cgconditions}
\omega A \ll 1 ~, \qquad \omega |\vec{x}|\ll 1~.
\end{equation}

When working in the linearized regime, it is convenient to rewrite the quark trajectory (\ref{xosc}) in terms of an exponential, $x^{\mu}=(t, Ae^{-i\omega t},0,0)$, with the understanding that henceforth one should in the end take the real part of any complex expressions. This notational change (and shift of phase from a sine to a cosine) will also facilitate the comparison with the results reported in \cite{cg}. Given our general result (\ref{trfsqpointlike}) for a pointlike quark (which, it is worth emphasizing, does not depend on the external force), the one element we need to obtain an explicit expression for the gluonic field is the  retarded time defined by the null interval condition $\tret=t-|\vec{x}-\vec{x}(\tret)|$. To linear order in $A$, it is easy to see that the solution is given by
\begin{equation}\label{tretosc}
\tret=t-|\vec{x}|+\frac{A x}{|\vec{x}|}e^{-i\omega(t-|\vec{x}|)}+\mathcal{O}(A^2)~.
\end{equation}

Using (\ref{tretosc}) in (\ref{trfsqpointlike}), and taking into account the first of the linearization conditions (\ref{cgconditions}),
the profile of the gluonic field is found to be
\begin{equation}\label{trfsosc}
\expec{\trFsq(x)}=\frac{\sqrt{\lambda }}{16 \pi ^2|\vec{x}|^4}+\frac{\sqrt{\lambda}(\vec{A}\cdot\vec{x} )}{4 \pi ^2|\vec{x}|^6}e^{-i\omega (t-|\vec{x}|)}(1-i\omega|\vec{x}|)+\mathcal{O}(A^2)~,
\end{equation}
where we have defined $\vec{A}\equiv (A,0,0)$.
As expected, the zeroth-order term in (\ref{trfsosc}) correctly reproduces the static contribution
\cite{dkk}. The term linear in $A$ represents the dynamical part of the field arising from the oscillatory motion of the quark, and, just like the full profile (\ref{trfsqpointlike}), manifestly contains a single time delay, corresponding to propagation strictly at the speed of light.

Let us now revisit the results of \cite{cg}. {}From
the linearized solution for the string embedding, the authors of that paper derived the static field contribution shown in the first term of (\ref{trfsosc}) (known previously from \cite{dkk})\footnote{It should be noted that Eq. (23) of \cite{cg} erroneously reports a denominator that is twice as large as that in (\ref{trfsosc}).}, and going beyond that, deduced the leading dynamical contribution to the gluonic field for small oscillation amplitude,
\begin{equation}\label{cgresult}
\expec{\trFsq(x)}^{\mbox{\scriptsize dyn}}_{\mbox{\scriptsize\cite{cg}}}
=\frac{\sqrt{\lambda}(\vec{A}\cdot\vec{x})}{32\pi^2}\int^{\infty}_0 dz' z'^2(1-i\omega z')f(\sqrt{z'^2+|\vec{x}|^2})e^{-i\omega (t-\sqrt{z'^2+|\vec{x}|^2}-z')}~,
\end{equation}
with
\begin{equation*}
f(u)=\frac{105}{u^9}-\frac{57i\omega}{u^8}-\frac{12\omega^2}{u^7}+\frac{i\omega^3}{u^6}~.
\end{equation*}
The combination in the integrand of the growing factor $z'^2(1-i\omega z')$ with the rapidly decaying $f(\sqrt{z'^2+|\vec{x}|^2})$ ensures that the main contribution to the integral is sharply localized around $z'\sim |\vec{x}|$, in accord with the UV/IR connection. In (\ref{cgresult}) we also see explicitly the advertised superposition of contributions with all possible time delays, seemingly in conflict with (\ref{trfsosc}).

To make an honest comparison between expressions (\ref{trfsosc}) and (\ref{cgresult}), we would need to carry out the integral in the latter. It is important to recall at this point that, even though this integral runs over all values of the bulk radial coordinate, the integrand is only accurate for $z'\ll 1/\omega$, because, as we explained before, the linearized approximation breaks down at large values of $z'$. The result of the integral can therefore only be trusted quantitatively at $|\vec{x}|\ll 1/\omega$, so in what follows we will confine ourselves to this observation region.  Along the lines of \cite{cg}, it is convenient to change the variable of integration in (\ref{cgresult}) to $\xi=\sqrt{1+z'^2/|\vec{x}|^2}+z'/|\vec{x}|$, to arrive at
\begin{equation}\label{cgresult2}
\expec{\trFsq(x)}^{\mbox{\scriptsize dyn}}_{\mbox{\scriptsize\cite{cg}}}
=\frac{\sqrt{\lambda}(\vec{x}\cdot\vec{A})}{8 \pi^2 |\vec{x}|^6}\int^{\infty}_1 \! d\xi\, h(\xi)e^{-i\omega(t-\xi |\vec{x}|)}
\end{equation}
with
\begin{equation*}
h(\xi)=\frac{\xi(\xi^2-1)^2}{(\xi^2+1)^5}
\!\!\left(2\xi-i\omega |\vec{x}|(\xi^2-1)\!\right)
\!\!\!\left[\frac{105 (2\xi)^3}{(\xi^2+1)^3}-\frac{57 i\omega|\vec{x}| (2\xi)^2}{(\xi^2+1)^2}-\frac{12 \omega^2 |\vec{x}|^2 (2\xi)}{(\xi^2+1)}+i\omega^3 |\vec{x}|^3\right].
\end{equation*}
This form emphasizes the presence of components with all possible propagation velocities $1/\xi$. The function  $h$ is a rational function of $\xi$ and a polynomial in the small quantity $\omega |\vec{x}|$, whose leading term is of order $(\omega |\vec{x}|)^0$. If we integrate by parts, differentiating the exponential and integrating $h(\xi)$, we are left with a surface term of this same order, which is evaluated at the lower endpoint $\xi=1$ (with the upper endpoint giving a vanishing contribution), and a remaining integral of order $(\omega |\vec{x}|)^1$:
\begin{eqnarray}\label{cgresult3}
\expec{\trFsq(x)}^{\mbox{\scriptsize dyn}}_{\mbox{\scriptsize\cite{cg}}}
&=&\frac{\sqrt{\lambda}(\vec{x}\cdot\vec{A})}{8 \pi^2 |\vec{x}|^6}
\left[{1\over 640}\left\{1280+15i(19\pi-64)\omega |\vec{x}|-1016(\omega |\vec{x}|)^2\right.\right.
\nonumber\\
{}&{}&\qquad\qquad\qquad\quad\;\left.+2i(128-5\pi)(\omega |\vec{x}|)^3+40(\omega |\vec{x}|)^4\right\} e^{-i\omega(t-|\vec{x}|)}
\nonumber\\
{}&{}&\qquad\qquad\quad\;\left.-i\omega |\vec{x}|\int^{\infty}_1 \! d\xi\, H(\xi)e^{-i\omega(t-\xi |\vec{x}|)}\right]~,
\end{eqnarray}
where $H$ of course denotes the primitive of $h$.

{}From (\ref{cgresult3}) we see that, to order $(\omega|\vec{x}|)^0$, the time-dependent gluonic field reduces to $\sqrt{\lambda}(\vec{x}\cdot\vec{A})/4 \pi^2 |\vec{x}|^6$, in precise agreement with (\ref{trfsosc}). We can continue integrating by parts iteratively to completely reexpress the result of \cite{cg} as a series in powers of the small parameter $\omega|\vec{x}|$. Already at order $(\omega|\vec{x}|)^1$ one finds disagreement with the exact small-amplitude result (\ref{trfsosc}), consistent with the fact that the integrand in (\ref{cgresult}) is not quantitatively reliable away from the small-distance regime (\ref{cgconditions}). In other words, carrying out the worldsheet integral and then linearizing, as we have done in this paper, does not yield the exact same answer as linearizing and then integrating, as was done in \cite{cg}. Nonetheless, what is clear is that, at any order, upon carrying out the integral in (\ref{cgresult}) one inevitably evaluates the result at the endpoint $\xi=1$, and is therefore left with a single phase factor $\exp[-i\omega(t-|\vec{x}|)]$.

The final conclusion is thus that the integrated expression (\ref{cgresult}), which manifestly displays the gluonic field as a sum of contributions with all possible time delays, actually gives the same result (within its range of validity) as the linearized and infinitely-massive version of our formula (\ref{trfsqfinal}) for the gluonic profile, which incorporates a single time delay. This proves that, despite appearances, there is in fact no contradiction between \cite{cg} and \cite{iancu1,iancu2}: it is just that the choice of worldsheet coordinates and the lack of an exact solution for the string embedding prevented the authors of \cite{cg} from being able to carry out the integral explicitly to find the \emph{net} retardation pattern.

\subsection{Uniform circular motion} \label{circularsubsec}

Let us now apply our general result (\ref{trfsqfinal}) to a quark undergoing uniform circular motion. This is another natural benchmark example, whose associated energy density distribution $\expec{T_{00}(x)}$ has been studied in detail for the case of an infinitely-massive quark \cite{liusynchrotron} (see also \cite{veronika}). Surprisingly, the radiation field was found to coincide with the synchrotron pattern familiar from classical electromagnetism, and therefore gave the first indication of the absence of temporal/radial broadening emphasized in \cite{iancu1,iancu2}. In this section we will consider the more general case of a quark with finite mass, and examine the gluonic field via the observable $\expec{\trFsq(x)}$, which we already know to be sensitive only to the near-field of the quark.

The trajectory of a quark moving with constant angular velocity $\omega$ along a circle of radius $\rho_m$   can be described by
\begin{equation}\label{rotatrajectory}
x^{\mu}=(t, \rho_m\cos{\omega t},\rho_m\sin{\omega t},0),
\end{equation}
with corresponding four-velocity
\begin{equation}\label{velocity}
v^{\mu}=(\gamma, -\gamma \omega \rho_m\sin{\omega t},\gamma\omega \rho_m\cos{\omega t},0)~,\quad \text{with}\quad \gamma=\frac{1}{\sqrt{1-\omega^2 \rho_m^2}}~.
\end{equation}
As in the previous section, to use our finite-mass formula (\ref{trfsqfinal}) expressed in terms of physical quark data via (\ref{atilde})-(\ref{stilde}), we would need first to solve the Lorentz-Dirac-like equation (\ref{lorentzdirac}) to determine  the external four-force acting on the quark.

Fortunately, for uniform circular motion there is a shortcut: it is easy to directly deduce the value of the auxiliary tilde variables (describing the motion of a fictitious string endpoint at the AdS boundary) that appear in (\ref{trfsqfinal}). The reason is that in this case the string embedding (\ref{mikhsoltilde}) or (\ref{mikhsolzm}) is known to take the form of a uniformly rotating spiral \cite{liusynchrotron}, and so, to satisfy the requirement (\ref{quarkx}) that its physical endpoint at $z_m$ undergo uniform circular motion, we merely need to require that the fictitious endpoint at $z=0$ also rotate uniformly (with an appropriately shifted phase). Since the radius of the spiral grows with the radial AdS coordinate according to \cite{liusynchrotron}
\begin{equation*}
\rho(z)=\rho_0\sqrt{1+\frac{\omega^2 z^2}{1-\omega^2\rho_0^2}}~,
\end{equation*}
we must choose the radius at the boundary, $\rho_0$, such that
\begin{equation}\label{rotatingradius}
\rho(z_m)=\rho_0\sqrt{1+\frac{\omega^2 z_m^2}{1-\omega^2\rho_0^2}}=\rho_m~.
\end{equation}

\begin{figure}[htb]
$$
\begin{array}{cc}
  \epsfig{width=2.5in,file=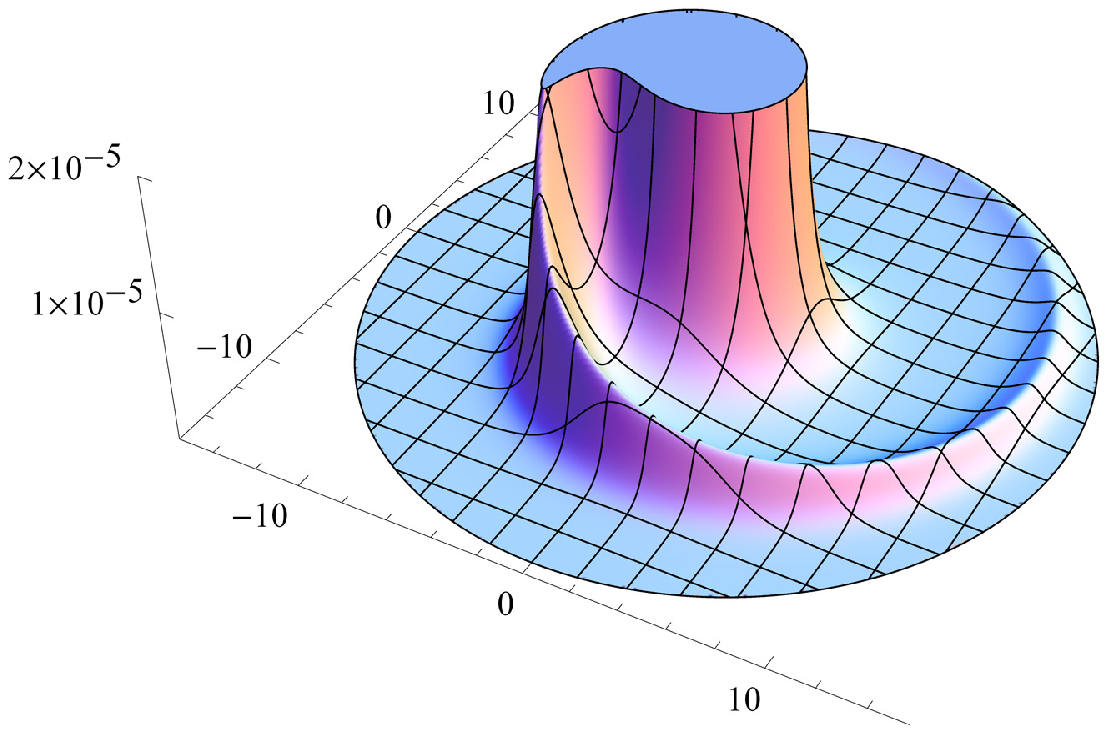} & \epsfig{width=2.5in,file=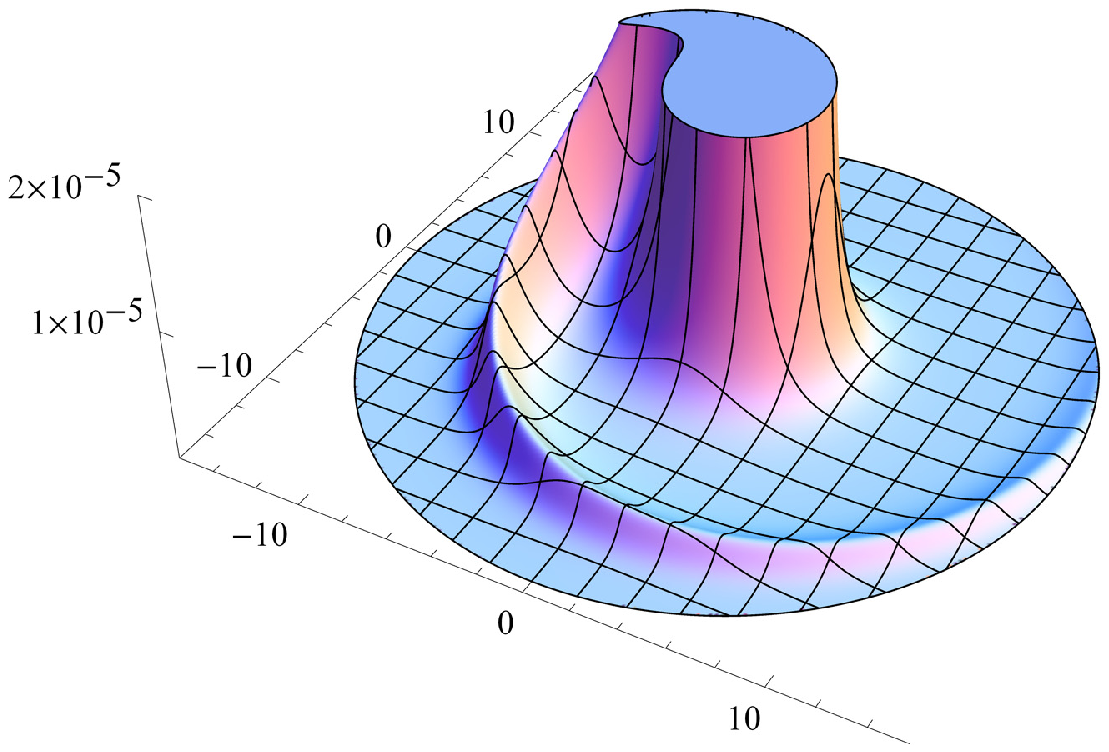} \\
  \epsfig{width=2.5in,file=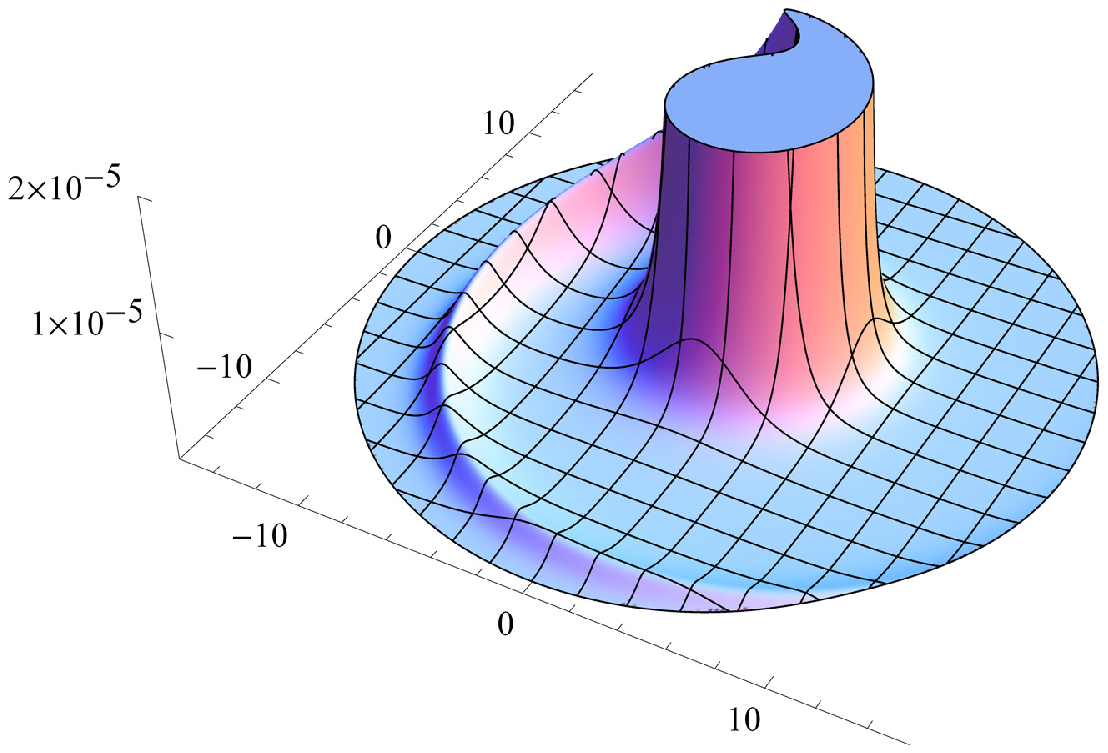} & \epsfig{width=2.5in,file=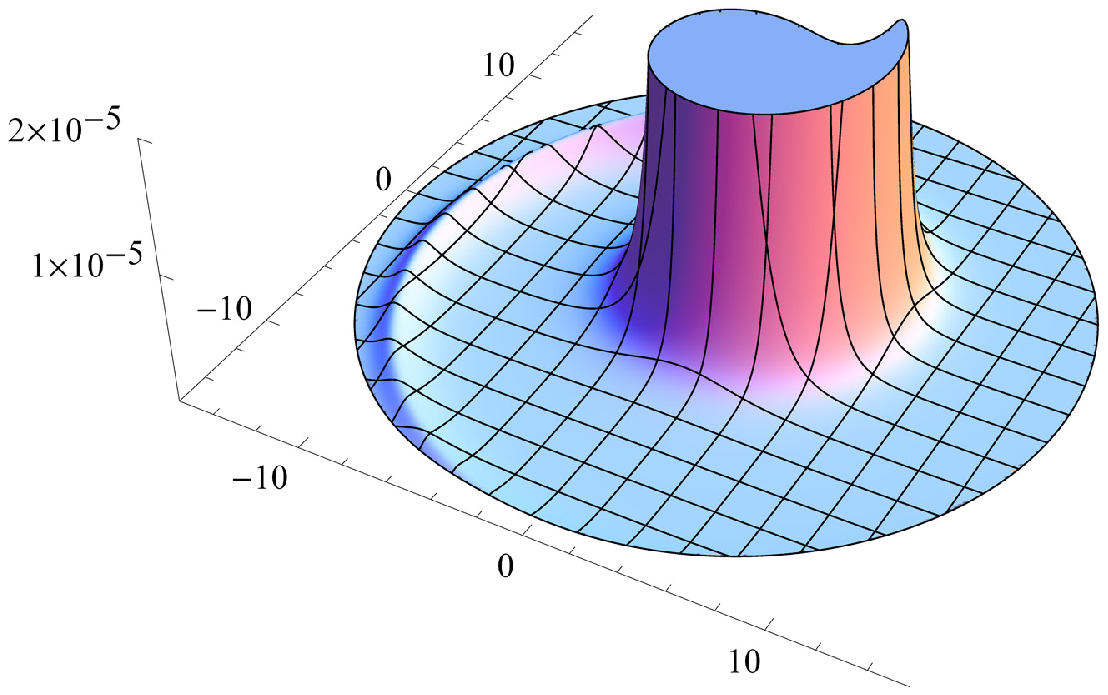}
\end{array}
$$
\begin{picture}(0,0)
 \put(14,78){\tiny $\expec{\tr F^2}$}
 \put(81,78){\tiny $\expec{\tr F^2}$}
 \put(14,36.5){\tiny $\expec{\tr F^2}$}
 \put(81,36.5){\tiny $\expec{\tr F^2}$}
 \put(94,73.5){\tiny $x$}
 \put(27,73.5){\tiny $x$}
 \put(94,32){\tiny $x$}
 \put(27,32){\tiny $x$}
 \put(101.5,49.5){\tiny $y$}
 \put(34.5,49.5){\tiny $y$}
 \put(101.5,8){\tiny $y$}
 \put(34.5,8){\tiny $y$}
 \end{picture}
 \vspace{-0.8cm}
\caption{\small Gluonic profile of a heavy quark undergoing uniform circular motion in the $xy$-plane, with parameters $z_m=1$, $\rho_m=1$ and $\omega=1/2$ (for which (\ref{rotatingradius}) yields $\rho_0$=0.874). The plots correspond to four successive snapshots of our simulation for $t=0,2,4$ and $6$ respectively (all within one period). The near-field waves display no broadening, and are clearly consistent with the radiation pattern observed in \cite{liusynchrotron,veronika,iancu2}.}
\end{figure}

To derive explicit results we additionally need to determine the retarded time $\tret$ by solving (\ref{tret}). Just like in the previous section, the equation turns out to be transcendental, so we process it numerically. The final result for $\expec{\trFsq(x)}$ is shown in Fig.~5, for a sample choice of parameter values. The spiral profile of the gluonic fields is clearly in close resemblance with the pattern detected in \cite{liusynchrotron,veronika,iancu1} through the observable $\expec{T_{00}(x)}$.  Indeed, one can check that as the quark of the mass approaches infinity, the pattern precisely reduces to the expected synchrotron form. At finite mass, we have a generalized synchrotron pattern, with somewhat different width of the spiral arm and a net subluminal propagation speed, as dictated by (\ref{tret}).   Still, no temporal broadening is observed and the disturbance remains sharply localized.\footnote{It would be interesting to follow the ideas presented for the energy density distribution in \cite{veronika} to try to recover these same results in terms of a superposition of dilatonic shock waves, but we leave this for future work \cite{bryan}.}

\subsection{Constant acceleration} \label{accelsubsec}

As a final example, let us consider a quark moving with constant \emph{proper} acceleration. The nature of the radiation emitted by a uniformly accelerated charge in the framework of classical electrodynamics is a topic that has received considerable attention over the years (see for instance \cite{accel} and the references therein). Various authors have already used holographic techniques to study this configuration in the previously uncharted, strongly-coupled regime \cite{accelembedding,brownianunruh,brownianunruh2}. In particular, it was shown that the physics behind the Unruh effect (on the field theory side) is nicely captured by the structure of the string embedding from the gravitational viewpoint.\footnote{The same mechanism was generalized recently to the case of an arbitrary accelerated trajectory on a bounded space \cite{mariano}.} The purpose of this subsection is to complement the previous analyses by showing explicitly the gluonic pattern sourced by such a quark.

\begin{figure}[htb]
\begin{center}
\epsfig{width=3in,file=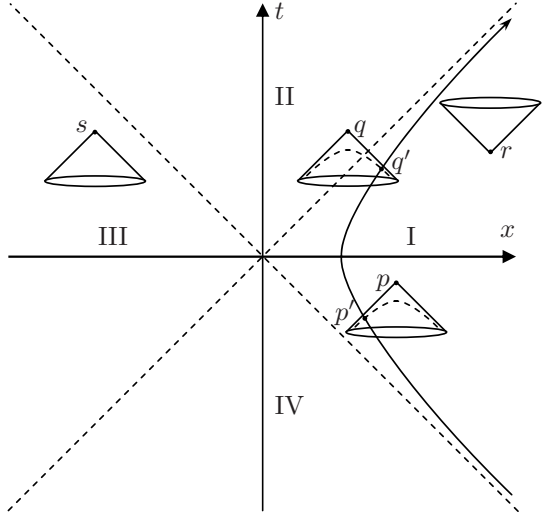}
\end{center}
\begin{picture}(0,0)
 \put(77.5,76.5){\footnotesize $t$}
 \put(107.5,47.5){\footnotesize $x$}
 \put(95,46.5){\footnotesize I}
 \put(77.5,65){\footnotesize II}
 \put(54,46.5){\footnotesize III}
 \put(77.5,24){\footnotesize IV}
 \put(91,41){\footnotesize $p$}
 \put(85.5,36.5){\footnotesize $p'$}
 \put(88,61.5){\footnotesize $q$}
 \put(93,56.5){\footnotesize $q'$}
 \put(107.5,58){\footnotesize $r$}
 \put(51,61.5){\footnotesize $s$}
 \end{picture}
\vspace{-1.2cm}
\caption{\small The hyperbolic trajectory $x^\mu(\tau)$ of a quark under uniformly accelerated
motion. The retarded time $\tret$ associated with any point is determined by the intersection between the quark worldline and the hyperbola (\ref{tau0}) contained inside the past light-cone of the given point. Only regions I and II are affected by the accelerated quark.}
\label{hyperb}
\end{figure}

Without loss of generality, we will assume that the motion is along the $x$-direction so, in covariant notation, the trajectory we are interested in is given by the usual hyperbola
\begin{equation}
x^\mu=(t,A^{-1}\sqrt{1+A^2t^2},0,0)~,\label{xmuacc}
\end{equation}where $A$ is the magnitude of the four acceleration,
\begin{equation} A\equiv \sqrt{{d^2 x_{\mu}\over d\tau^2}{d^2 x^{\mu}\over d\tau^2}}={d\over dt}(\gamma
v)~,
\end{equation}and where for convenience we have made a particular choice of the spacetime origin. From (\ref{xmuacc}) it follows that
\begin{equation}
\gamma(t)=\sqrt{1+A^2t^2}~,
\end{equation}
and
\begin{equation}
\tau=A^{-1}\mathrm{arcsinh}(At)~,
\end{equation}
where $\tau$ denotes, as usual, the time measured by the internal clock of the accelerated observer. The worldline described by such a hyperbola is displayed in Fig.~\ref{hyperb}.

Again, determining the gluonic profile in the general case (\ref{trfsqfinal}) requires knowledge of not only the quark trajectory and its derivatives, but also the total force applied to it. In this case, the equation of motion (\ref{lorentzdirac}) reduces simply to
\begin{equation}\label{ldacce}
A(1-z_m^4\bar{F}^2)=z_m\bar{F}\sqrt{1-z_m^4\bar{F}^2}+z_m^2\dot{\bar{F}}\sqrt{1+A^2t^2}~.
\end{equation}
Notice that, just as in the examples that we have studied before, the equation obtained is a non-linear first order differential equation, and this automatically implies that there are an infinite number of solutions depending on the initial condition. Numerical integration of (\ref{ldacce}) reveals a structure similar the one found in Section~\ref{oscillatesubsec}: for generic initial conditions, the solution is a non-linear superposition of a transient response that decays rapidly (within a time scale of order $z_m$) and a forced (steady-state) solution that dominates at late times and is independent of the initial condition. This is presumably an inherent property of the structure of the generalized Lorentz-Dirac equation (\ref{lorentzdirac}), associated with the presence of radiation damping, and, given the acquired numerical evidence, we expect it to hold also for more general/arbitrary trajectories.

As noted in Section~\ref{oscillatesubsec}, the choice of initial condition for the force encodes the past behavior of the quark, so if we wish to describe the situation where the quark has been undergoing uniform acceleration for all times, we must effectively ignore the transient tail. Equivalently, we can give the initial condition for the force at a finite time, but tune it to coincide with the asymptotic, steady-state, value. For arbitrary $A$, this value is seen from (\ref{ldacce}) to be
\begin{equation}\label{facce}
\bar{F}=\frac{A}{z_m\sqrt{1+A^2z_m^2}}~.
\end{equation}

If our interest lies solely on the perpetually accelerating quark, we could have in fact avoided any mention of the Lorentz-Dirac-like equation (\ref{ldacce}), because in this case the string embedding (\ref{mikhsoltilde}) or (\ref{mikhsolzm}) is known to take the simple form $X(t,z)=\sqrt{A^{-2}+t^2-z^2+z_m^2}$ \cite{accelembedding,brownianunruh}, which makes it easy to directly deduce the behavior of the auxiliary tilde variables required in (\ref{trfsqfinal}): uniform acceleration at the physical string endpoint $z=z_m$ corresponds to uniform acceleration (of different magnitude) at the fictitious endpoint at $z=0$. We should stress that this simple relation between the behavior of the string embedding at different radial depths, which we also encountered for the case of uniform circular motion in Section~\ref{circularsubsec}, is not by any means generic, and in particular does not hold for the oscillating quark studied in Section~\ref{oscillatesubsec} (for which harmonic oscillation of the physical endpoint is easily seen to translate into nonharmonic oscillation of the fictitious boundary endpoint).

To determine the gluonic profile through (\ref{trfsqfinal}), we additionally need to solve (\ref{tau0}) to obtain the retarded time $\tau_0$, and then evaluate all the data at this time. For convenience we will use its noncovariant equivalent (\ref{tret}), which can be inverted explicitly to deduce that
\begin{eqnarray}\label{t0acce}
\tret=&&\!\!\!\!\!\!\!\frac{1}{2 A^2(x^2-t^2)(1-z_m^2A^2)}\bigg[\left(1+A^2 \left(\vec{x}^{\,2}-t^2\right)\right) (t-z_m A x)-\\
&&\!\!\!\!\!\!\!\sqrt{(x-z_m A t)^2 \left(1+2 A^2 \left(t^2-2x^2+\vec{x}^{\,2}\right)+A^4 \left(\left(\vec{x}^{\,2}-t^2\right)^2+4 \left(x^2-t^2\right) z_m^2\right)\right)}\bigg].\nn
\end{eqnarray}

This can be interpreted geometrically as follows. The retarded time $\tret$ associated with a given point $(t,x,y,z)$ corresponds to the (unique)
intersection of the hyperbola (\ref{tret}), contained inside the \emph{past} light-cone of $(t,x,y,z)$, with the quark worldline $x^{\mu}(\tau)$. For
instance, in Fig.~\ref{hyperb}, the points $p'$ and $q'$ define the retarded times
$t^p_{\mbox{\scriptsize ret}}$ and $t^q_{\mbox{\scriptsize ret}}$ associated with the points $p$ and $q$, respectively.  The fact that the signal travels along the hyperbola (\ref{tret}) instead of along a null line means that the velocity of propagation is subluminal; we will come back to this point in the next section. For $z_m\rightarrow0$ the hyperbola converges to the lightcone, and the retarded time is determined by the point of intersection between the quark worldline and the past lightcone. In this case the signal travels strictly at the speed of light. Note also that only regions I and II are affected by the
fields due to a charged particle with a worldline given by (\ref{xmuacc}). On the contrary, points in regions III and IV, are causally disconnected from the quark worldline (see for example the point $s$ in Fig. \ref{hyperb}) and this automatically implies that the gluonic field vanishes there.
\begin{figure}[htb]
$$
\begin{array}{cc}
  \epsfig{width=2.5in,file=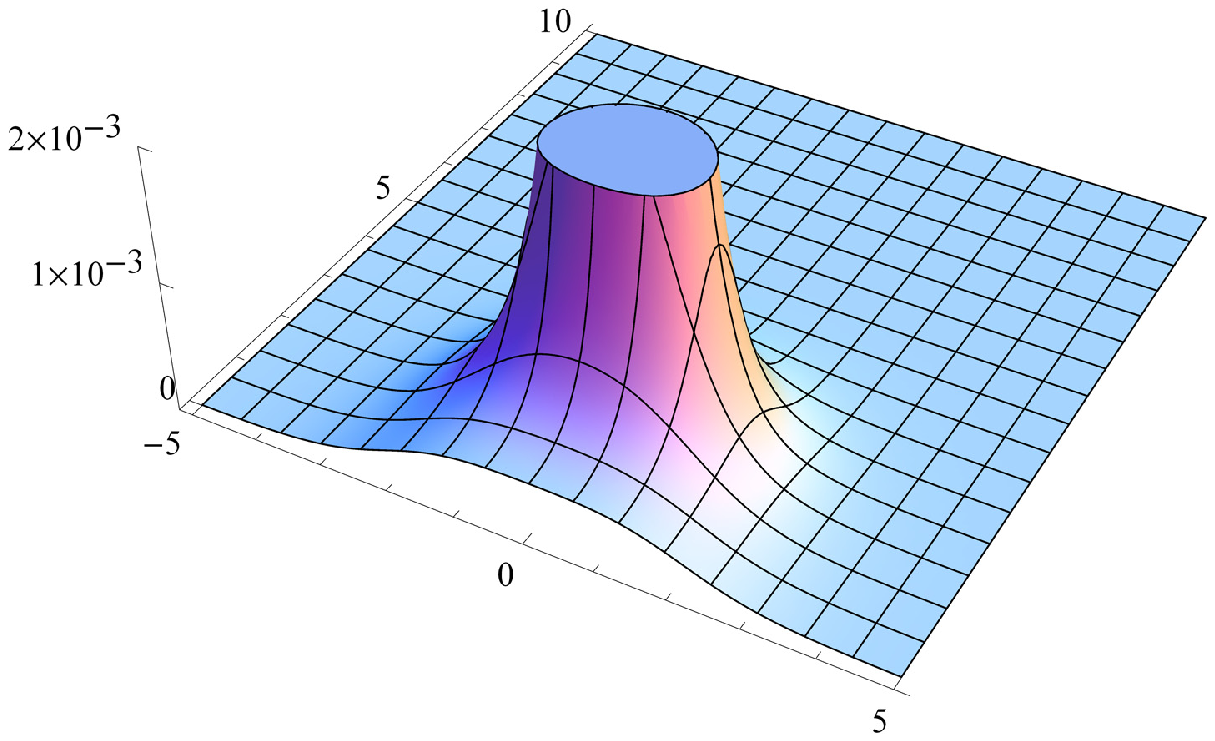} & \epsfig{width=2.5in,file=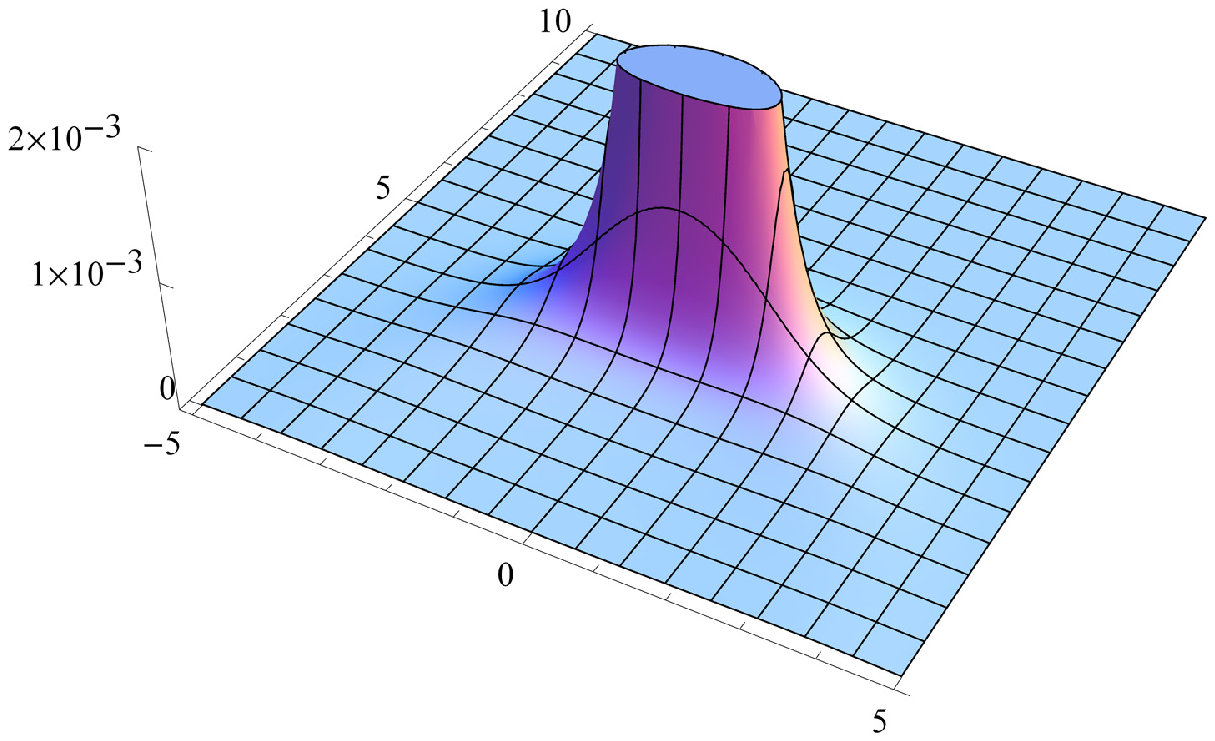} \\
  \epsfig{width=2.5in,file=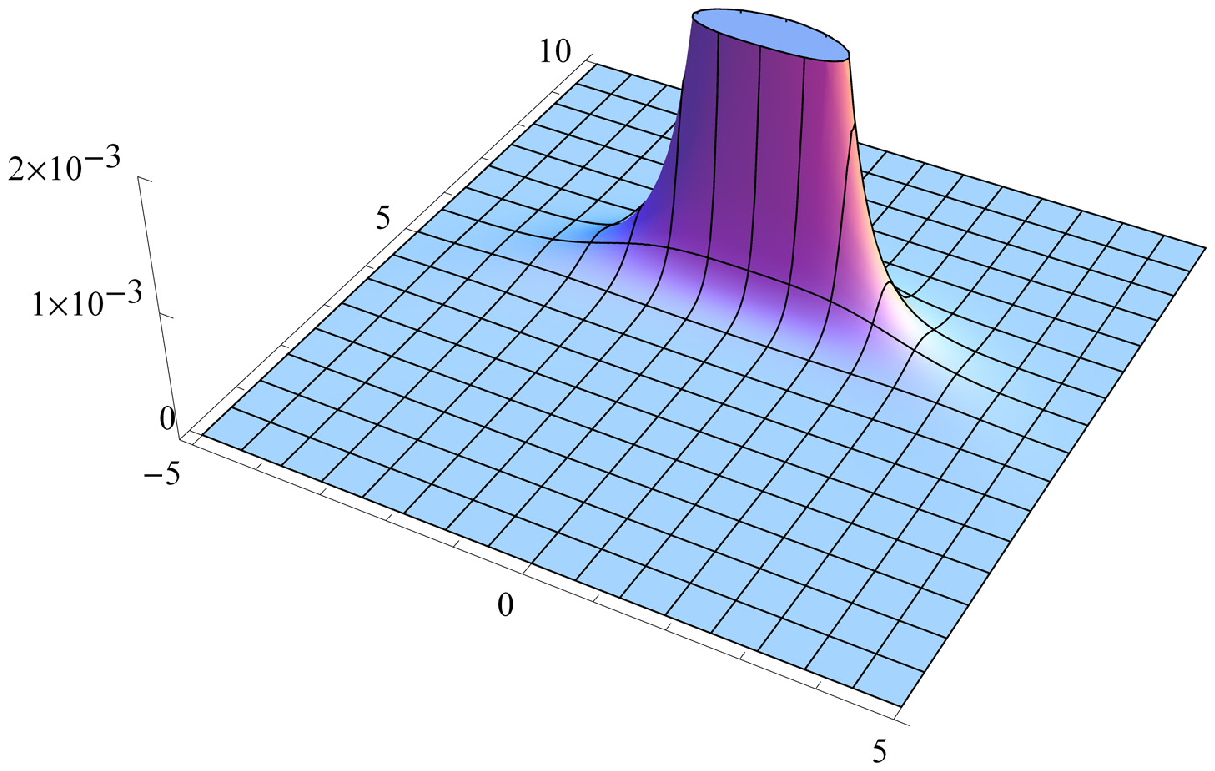} & \epsfig{width=2.5in,file=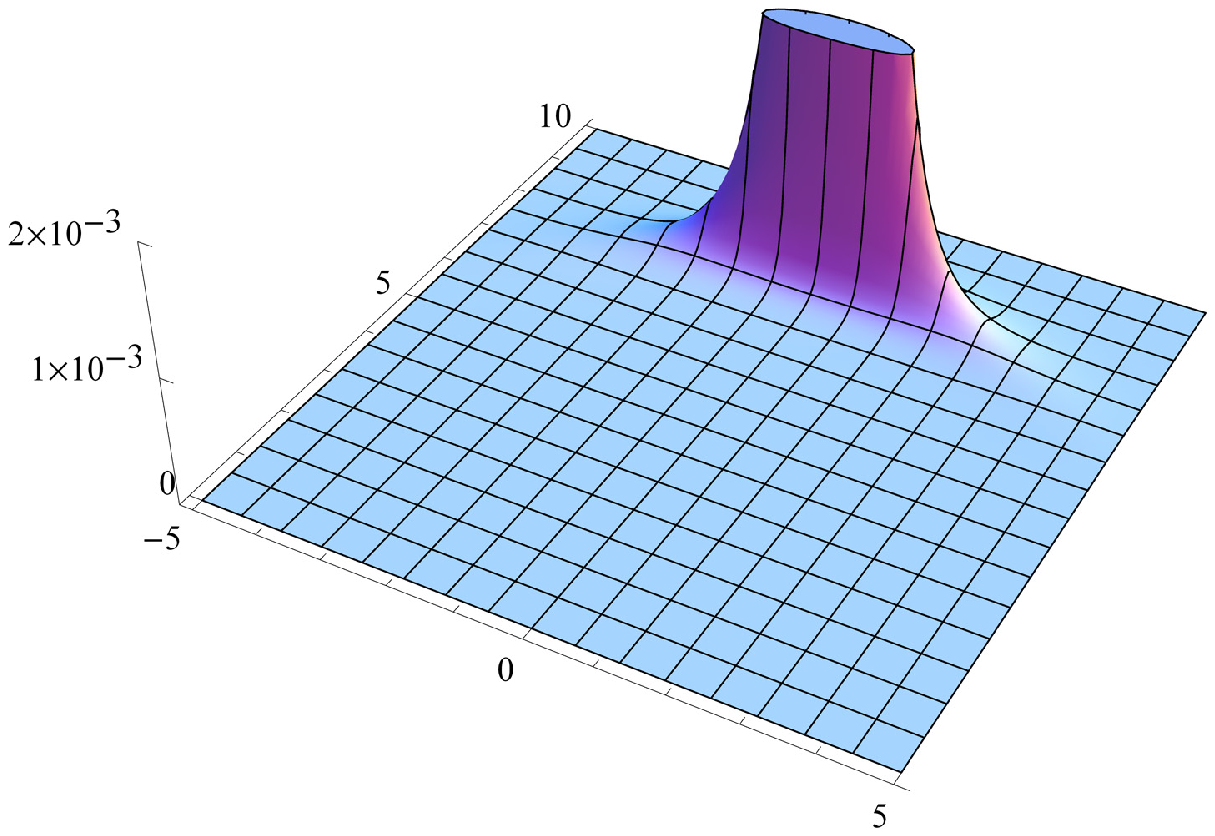}
\end{array}
$$
\begin{picture}(0,0)
 \put(14,84){\tiny $\expec{\tr F^2}$}
 \put(81,84){\tiny $\expec{\tr F^2}$}
 \put(14,38.5){\tiny $\expec{\tr F^2}$}
 \put(81,38.5){\tiny $\expec{\tr F^2}$}
 \put(94,79.5){\tiny $x$}
 \put(27,79.5){\tiny $x$}
 \put(94,33.5){\tiny $x$}
 \put(27,33.5){\tiny $x$}
 \put(101.5,55.5){\tiny $y$}
 \put(34.5,55.5){\tiny $y$}
 \put(101.5,9.5){\tiny $y$}
 \put(34.5,9.5){\tiny $y$}
 \end{picture}
\vspace{-0.8cm}
\caption{\small Gluonic profile of a heavy quark with uniform proper acceleration in the $x$-direction, with the horizontal axes in units of the quark Compton wavelength, $z_m=1$, and the vertical axis in units of $\sqrt{\lambda}$. The plots correspond to four successive snapshots of our simulation, for $t=2,4,6$ and $8$, respectively, and acceleration $A=1/2$.  The value of the gluonic profile at the location of the quark is finite but this section is capped off to show the near field details. The pattern obtained displays the expected Lorentz contraction in the transversal direction as the quark increases its velocity.}\label{figacce}
\end{figure}

Finally, using (\ref{xmuacc}) and (\ref{facce}) and evaluating (\ref{trfsqfinal}) in (\ref{t0acce}) we obtain the gluonic field profile shown in Fig.~\ref{figacce}. We choose sample values $z_m=1$, $\sqrt{\lambda}=1$, $A=1/2$ and we plot points in (part of) regions I and II. The figures correspond to four successive snapshots of the profile, for $t=2,4,6$ and $8$ respectively. The most salient difference of the finite mass case with respect to the $z_m\rightarrow 0$ case is that the extended nature of the quark makes the gluonic profile finite even at the location of the quark. The overall pattern decays very fast however, and for points very far from the quark there is no significant difference between these two cases. The behavior of the gluonic field displays the expected Lorentz contraction in the longitudinal direction as the quark increases its velocity.

It would be interesting to investigate this setup further through the calculation of $\expec{T_{\mu\nu}(x)}$ for both the inertial observer and the comoving observer, but this computation lies beyond the scope of this paper. The radiation field (i.e., the part that decays as $1/|\vec{x}|^2$) could be extracted from that observable, and contrasted with the expected result from classical electrodynamics \cite{accel} (with which contact has already been made for the energy density in the case of an infinitely massive quark \cite{iancu2}). It would also be interesting to exploit the tensorial properties of our result as in \cite{gupta,brownianunruh} to obtain the gluonic profile as seen by the comoving observer.

\section{Discussion}\label{discussionsec}

Let us now discuss the physical implications of our results. According to (\ref{trfsqfinal}), data of the quark (position, velocity, acceleration and jerk, together with external force, yank and tug) at the sole instant (\ref{tret}) on its trajectory determine the expectation value of $\trFsq(x)$. It is important to remember that, just as in the equation of motion (\ref{lorentzdirac}), the presence of the external force in our final expression for the one-point function conveniently replaces a dependence on an infinite number of higher-derivative terms \cite{damping}, and is therefore indicative of nonlocality. In our result, these higher derivatives appear modulated by powers of the quark Compton wavelength $z_m$, so, in essence, it is data within $z_m$ of $\tret$ that have an impact on the gluonic field at the chosen observation point. The nonlocality is thus naturally associated with the extended nature of the quark.

This nonlocality disappears in the pointlike/infinitely-massive limit $z_m\to 0$, where, as noted at the end of Section \ref{arbitrarysec}, our general result reduces to (\ref{trfsqpointlike}), which depends only on the position and velocity of the quark at a retarded time evaluated on the past lightcone of the observation point, just as in classical electrodynamics.
As a matter of fact, our result (\ref{trfsqpointlike}) for an infinitely massive quark can easily be seen to agree with the square of the Lienard-Wiechert electromagnetic field strength (more precisely, $(1/4)F^2$) set up classically by a pointlike electric charge $e$ \cite{jackson}, under the replacement\footnote{Notice that this is somewhat different from the replacement $e^2\to 3\sqrt{\lambda}/4\pi$ that translates the radiated energy in classical electromagnetism to the one in strongly-coupled SYM \cite{mikhailov}.} $e^2\to\sqrt{\lambda}/8\pi^2$.
Equivalently, it agrees with the leading order result within $\cN=4$ SYM at \emph{weak} coupling, with the rescaling $\lambda\to 4\sqrt{\lambda}$, which is exactly the same connection found in \cite{iancu2} for the expectation value of the energy density. As in electromagnetism, the gluonic field (\ref{trfsqpointlike}) is simply the boosted version of the Coulombic profile  set up by a static pointlike quark  \cite{dkk}, at the appropriate retarded time.

In the pointlike ($z_m=0$) case, then, the retardation pattern found for the gluonic field observable under present examination, $\expec{\trFsq(x)}$, is seen to be exactly the same as that obtained for the energy density, $\expec{T_{00}(x)}$, in \cite{liusynchrotron} and, especially, \cite{iancu1,iancu2} (as well as for other observables in \cite{iancu1}): the net profile propagates strictly at the speed of light, with no temporal (or, equivalently, radial) broadening.

In the case of a non-pointlike/finitely-massive ($z_m>0$) quark (which was not examined in \cite{liusynchrotron,iancu1,iancu2}), we have found that two features of the gluonic profile are modified. First, signals no longer travel at the speed of light, but rather propagate along the timelike interval (\ref{tau0}). The average velocity of propagation of the signal, $\vec{v}_{\mbox{\scriptsize average}}\equiv (\vec{x}-\vec{x}(\tret))/(t-\tret)$ follows from (\ref{tret}) as
\begin{equation}\label{vsignal}
v_{\mbox{\scriptsize average}}=\sqrt{1-\frac{z_m^2}{(t-\tret)^2}}~.
\end{equation}
Since $z_m\le t-\tret <\infty$, (\ref{vsignal}) implies that $0\le v_{\mbox{\scriptsize average}}<1$. {}From the gravity perspective, (\ref{tret}) clearly just means that the net signal from the string to the observation point on the AdS boundary has the smallest possible delay: the time for null propagation starting at the string endpoint (as opposed to any other location on the string). The interpretation of this delay from the gauge theory perspective is more interesting: it is due to the finite size of the dressed quark. Consider the seemingly most peculiar case where $v_{\mbox{\scriptsize average}}=0$, which happens when $t-\tret=z_m$. In this case $\vec{x}-\vec{x}(\tret)=0$, so the spatial location of the quark at the relevant instant coincides with that of the observation event, and it might seem surprising that there is still a delay before the gluonic field `takes notice'. Recalling, however, that $z_m$ is nothing but the size of the quark, we see that this result makes perfect physical sense: the delay just indicates the time for information to propagate from the `edge' of the quark to the observation point located at its (instantaneous) center. We can see in (\ref{tret}) that this same delay is present in the more general case where  $v_{\mbox{\scriptsize average}}>0$, but becomes progressively less noticeable as the distance between the quark and the observation point increases.

We emphasize that the full pattern of propagation is very complicated, with each point on the non-Abelian medium potentially yielding a contribution to the gluonic field at the observation point, and $v_{\mbox{\scriptsize average}}$ is associated solely with the delay in the net result. At first sight, it might seem odd that this velocity is not fixed, but this is in fact unavoidable: for subluminal propagation, the only Lorentz-covariant possibility is for signals to be emitted along hyperboloids at constant timelike interval, precisely as prescribed by (\ref{tau0}).

 A second modification in the non-pointlike profile (\ref{trfsqfinal}) with respect to (\ref{trfsqpointlike}) is that, as we mentioned at the beginning of this section, it depends on more data than just the position and velocity of the quark. In particular, the acceleration, jerk and snap of the quark have an impact on the gluonic field. Given the spatial falloff ($\propto 1/|\vec{x}|^4$) of $\expec{\trFsq(x)}$, these higher-derivative contributions cannot be interpreted as indicative of the presence of radiation. Of course, radiation \emph{is} expected to be generally present when the quark accelerates, but its unambiguous detection requires a generalization of the calculation in $\expec{T_{00}(x)}$ to the case of finite mass. The higher-derivative terms  in our result for $\expec{\trFsq(x)}$ are due to (or rather, evidence) the fact that the gluonic cloud of the quark is deformed when the latter accelerates (or jerks or snaps) \cite{damping}.

In spite of these differences, one important feature shared by the pointlike and non-pointlike results is that they are determined by the behavior of the quark/endpoint at a single retarded time $\tau_0$ or $\tret$.
It is worth recapping here how this result came about.
At the start of the computation, we had the expression (\ref{dilsol}) for the dilaton field in the bulk, which is \emph{a priori} sourced by all points on the string worldsheet. As we emphasized below (\ref{Uminmax}) and illustrated in Figs.~2 and 3, the causal structure of the propagator (\ref{prop}) appropriately restricts the source points to lie within a specific swath on the worldsheet, which covers a \emph{finite} temporal extent (parametrized by $t'$, $t'_r$ or $\tau'$) at any given radial depth in the bulk (parametrized by $z'$ or $U$). As we also noted, this swath shrinks to a line when, in accordance with the GKPW recipe \cite{gkpw}, the observation point is taken towards the AdS boundary. The requirement of propagation all the way to $z\to 0$ thus picks out only one instant on the worldsheet at each radial depth, and the dilaton/gluonic field profile at a given (Minkowski) spacetime location is obtained as the integral over all such contributions. The final, crucial ingredient in our calculation was the fact that, as seen in (\ref{dilsolmikh2}), the integrand in this radial integral, when expressed in terms of the $(\tau,U)$ parametrization, turned out to be a \emph{total derivative}, allowing the result to be written purely as a surface term. This is why, when all the dust settles, the observed $\expec{\trFsq(x)}$ depends only on the behavior of the (lower) string endpoint (with the upper endpoint excluded by causality), and thus involves a single retarded time.

{}In the calculation, our being left with a total derivative in (\ref{dilsolmikh2}) can be traced back to the cancelation of the $z'^2$ terms in (\ref{Umikh}). An analogous cancelation was seen to take place in the computation of the energy density \cite{iancu2}. Its appearance here in the midst of a scalar calculation shows that the effect is not specific to the interplay between the various tensor components relevant to \cite{iancu2}. In our setting, the cancelation clearly stems from the specific structure of the string embedding (\ref{mikhsol}) or (\ref{mikhsolzm}).  Physically, then, the absence of temporal broadening in our result is closely associated with the use of a purely retarded solution. The same feature is seen when using a purely advanced configuration, but would not be expected
in the more general case involving nonlinear superpositions of retarded and advanced solutions (or in the presence of a thermal medium \cite{gluonicprofile}, which, after an initial period, naturally brings into play analogous superpositions \cite{dragtime,dampingtemp}).

Given that we are dealing with a strongly-coupled non-Abelian gauge theory, it is remarkable that we have obtained a gluonic field profile that can be expressed in the relatively simple form (\ref{trfsqfinal}) (or, in the pointlike quark limit, (\ref{trfsqpointlike})), which depends on a single retarded time (\ref{tau0}). Since the Yang-Mills field is a nonlinear medium, we would naturally expect it to generate infinitely broadened disturbances, arising from rescattering  at all possible length/energy scales of the original signal emitted by the quark, leading to a pattern such as the one reported in \cite{cg}, with components propagating at speeds arbitrarily slower than light. Indeed, the authors of \cite{iancu1,iancu2} have argued that the absence of temporal broadening in results such as (\ref{trfsqpointlike}) is a physically unwanted feature, and indicates a failure of the supergravity approximation of AdS/CFT to capture the full quantum dynamics.

In our computation, however, we have seen how the final result (\ref{trfsqfinal}) or (\ref{trfsqpointlike}) \emph{is} in fact assembled from contributions of \emph{all} points along the string, which, via the UV/IR connection \cite{uvir}, translate precisely into the expected contributions from the SYM fields at all possible length scales. The appearance of a total derivative in (\ref{dilsolmikh2}) enabled us to explicitly express the result of the radial integral as a boundary term, but this does not imply that all the contributions from interior points of the string are canceling out,
only that their sum can be rewritten purely as a (different) function of  data at the string endpoint. This property would have been much more difficult to notice if we had chosen to carry out the analysis in the original static gauge $(t',z')$, or, more generally, in a parametrization not closely related our geometric choice of the retarded proper time $\tau'$ and the invariant AdS distance $U$. In such cases, we would be left contemplating a sum over contributions that manifestly involve time delays corresponding to all possible subluminal speeds, even though, of course, the final result of the sum is reparametrization invariant and would therefore necessarily coincide with what we have found here. As we have seen very explicitly in Section \ref{oscillatesubsec}, the study of an oscillating quark in \cite{cg} provides us with a pre-existing example of precisely this sort of situation.

Our physical interpretation is thus different from that of \cite{iancu1,iancu2}, even though the properties of our results are in full agreement with theirs. We see the AdS calculation yielding a one-point function that is manifestly obtained as a sum over contributions of the gluonic field at all possible length scales, just as one would expect, and therefore take (\ref{trfsqfinal}) and (\ref{trfsqpointlike}) at face value, as a \emph{prediction} of the AdS/CFT correspondence for the spacetime pattern emerging in the $\lambda,N_c\to\infty$ limit where the supergravity calculation is performed, for the specific case where the field configuration is assumed to be purely outgoing.
As we mentioned in the Introduction, at least in retrospect it feels to some extent natural for this assumption to imply a restriction on the overall retardation pattern. What is certainly surprising is the actual form of the predicted restriction: even though we would \emph{a priori} expect the net result to depend on data characterizing the entire quark trajectory, our final pointlike profile (\ref{trfsqpointlike})  is found to be completely local, and its non-pointlike extension (\ref{trfsqfinal}) is seen to have a nonlocality that (being bounded by the quark Compton wavelength (\ref{zm})) is directly associated with the extended nature of the finitely-massive quark, rather than with the nonlinear nature of the SYM fields.

There are several directions along which this circle of ideas might be profitably extended. It would certainly be interesting to inquire to what extent the surprising pattern of unbroadened propagation obtained in \cite{iancu1,iancu2} and the present paper is present in other setups made available to us by gauge/gravity duality (e.g., in nonconformal field theories). Another such direction, which we hope to report on in the near future \cite{juanfelipe}, is to revisit the calculation in \cite{iancu2} of the energy density in the gluonic field generated by a quark undergoing arbitrary motion, and generalize it to the case of finite quark mass. Taken together with the results of the present paper, this could shed light on the appropriate spacetime separation of the near- and radiation- field contributions, examined previously from a worldsheet perspective in \cite{dragtime,lorentzdirac,damping}, and also touched upon in \cite{iancu2}. Through these and a myriad other ongoing explorations, the correspondence still seems poised to teach us many new and important lessons about the dynamics of strongly-coupled gauge theories.

\section*{Acknowledgements}
We are grateful to Elena C\'aceres, Tomeu Fiol, Veronika Hubeny, David Mateos and \'Angel Paredes for useful discussions and comments on the manuscript. The present work was partially supported by Mexico's National Council of Science and Technology (CONACyT) grant 104649.
The research of M.C. is supported by 2009-SGR-168, MEC FPA2010-20807-C02-01, MEC FPA2010-20807-C02-02, CPAN CSD2007-00042 Consolider-Ingenio 2010. J.F.P. would like to thank the Texas Cosmology Center for partial support.

\appendix

\section{Appendix}

Here we give some more detail on the calculation of the one-point function of $\trFsq$ for an arbitrary quark trajectory presented in Section \ref{arbitrarysec}.
It helps to summarize the results (\ref{ttautaylor}), (\ref{ttau0}), (\ref{ttaucoeffs}) by writing
\begin{equation}\label{ttautaylor2}
\ttau=\ttau_0+A_1\Umin z + (A_2+A_3 \Uminsq)z^2+ \cO(z^3)~,
\end{equation}
where the coefficients $A_j$ are  independent of $\Umin$ and $z$, and are given explicitly  by
\begin{eqnarray}\label{As}
A_1&=&-\frac{z_m}{z_m+(x-\tx')\cdot(\tv'+\ta' z_m)}~,\nonumber\\
A_2&=&\frac{1}{2\left[z_m+(x-\tx')\cdot(\tv'+\ta' z_m)\right]}~,\\
A_3&=&-\frac{z_m^2\left(1+(x-\tx')\cdot(\ta'+\tj' z_m)\right)}{2\left[z_m+(x-\tx')\cdot(\tv'+\ta' z_m)\right]^3}~. \nonumber
\end{eqnarray}

Straightforward algebra then shows that
\begin{equation}\label{denom}
 (x-\tx(\ttau'))^2+z^2=B_1+B_2\Umin z + (B_3 + B_4\Uminsq)z^2+ \cO(z^3)~,
\end{equation}
with
\begin{eqnarray}\label{Bs}
B_1&=& (x-\tx')^2~,\nonumber\\
B_2&=&-2A_1(x-\tx')\cdot\tv'~,\\
B_3&=&1-2A_2 (x-\tx')\cdot\tv'~, \nonumber\\
B_4&=&-2A_3(x-\tx')\cdot\tv'-A_1^2(1+(x-\tx')\cdot\ta')~;\nonumber
\end{eqnarray}
whereas
\begin{equation}\label{jacobian}
z_m+(x-\tx(\ttau'))\cdot(\tv(\ttau')+ z_m\ta(\ttau'))=C_1+C_2\Umin z + (C_3 + C_4\Uminsq)z^2+ \cO(z^3)~,
\end{equation}
with
\begin{eqnarray}\label{Cs}
C_1&=& z_m+(x-\tx')\cdot(\tv'+\ta' z_m)~,\nonumber\\
C_2&=&A_1\left(1+(x-\tx')\cdot(\ta'+\tj' z_m)\right)~,\\
C_3&=&A_2 \left(1+(x-\tx')\cdot(\ta'+\tj' z_m)\right)~, \nonumber\\
C_4&=&A_3\left(1+(x-\tx')\cdot(\ta'+\tj' z_m)\right)
+{1\over 2}A_1^2\left((x-\tx')\cdot(\tj'+\ts' z_m)+z_m\ta'^2\right)~.\nonumber
\end{eqnarray}

Combining (\ref{dUmin}), (\ref{denom}) and (\ref{jacobian}) and expanding $ d\ttau'\, z/ ((x-\tx(\ttau'))^2+z^2)$, we find terms of order $z^2$ and $z^3$ that have up to a linear dependence on $\Umin$, and therefore vanish upon carrying out the integral (\ref{dilsolmikh3}). The leading contribution to the dilaton field in the $z\to 0$ limit arises from the terms of order $z^4 \Uminsq$, which yield
\begin{equation}\label{dilsolmikh4}
\varphi(x,z)=  {\sqrt{\lambda}z_m\,z^4\over 32\pi^{2}}
 \left( \frac{B_1^2(C_2^2-C_1 C_4)+ B_1 B_2 C_1 C_2 + (B_2^2-B_1 B_4) C_1^2  }{B_1^3 C_1^3}\right)+\cO(z^5)~.
\end{equation}
Plugging this into (\ref{trfsq}) and using (\ref{Bs}) and (\ref{Cs}), we finally arrive at the result (\ref{trfsqfinal}).

\end{document}